\documentclass{aa}  
\usepackage{graphicx}
\usepackage{txfonts}
\usepackage{hyperref}
\usepackage{natbib}
\usepackage{xcolor}
\usepackage{textgreek}
\usepackage{lipsum}
\usepackage{lscape}

\newcommand{\msol}{M_\odot}
\newcommand{\mpa}{\,M_\odot\mathrm{/yr}}
\newcommand{\rsol}{R_\odot}
\newcommand{\K}{\,\mathrm{K}}

\newcommand{\days}{\,\mathrm{d}}
\newcommand{\ltt}{\log \tau_\mathrm{KH}/\tau_{\dot M}}
\newcommand{\Mzams}{M_\mathrm{i}}

\let\eps\varepsilon

\begin{document} 

\title{Exploring the borderline between stable mass transfer and mergers in close binary evolution}

\author{C.~Schürmann\inst{1}\fnmsep\thanks{email: \texttt{chr-schuermann@uni-bonn.de}}
\and N.~Langer\inst{1}\fnmsep\inst{2}
}

\institute{Argelander-Institut für Astronomie, Universität Bonn, Auf dem Hügel 71, 53121 Bonn, Germany
\and Max-Planck-Institut für Radioastronomie, Auf dem Hügel 69, 53121 Bonn, Germany
}

\authorrunning{C. Schürmann \& N. Langer}
\date{Submitted ??? / Accepted ???}

\abstract{The majority of massive stars reside in binary systems, which are expected to experience mass transfer during their evolution. However, so far the conditions under which mass transfer leads to a common envelope, and thus possibly to a merging of both stars, are not well understood. Main uncertainties arise from the possible swelling of the mass gainer, and from angular momentum loss from the binary system, during non-conservative mass transfer. We have computed a dense grid of detailed models of stars accreting mass at constant rates, to determine their radius increase due to their thermal disequilibrium. While we find that models with faster than thermal timescale accretion generally expand, this expansion remains quite limited in the intermediate mass regime even for accretion rates which exceed the thermal timescale accretion rate by a factor of 100. Our models of massive accretion stars expand to extreme radii under those conditions. When the accretion rate exceed the Eddington accretion rate, our models expand dynamically. We have derived analytical fits to the radius evolution of our models and a prescription for the borderline between stable mass transfer and mergers for arbitrary accretion efficiencies. We then apply our results to grids of binary models adopting various constant mass transfer efficiencies and angular momentum budgets. We find that the former parameter has the stronger effect on the outcome of the Roche lobe overflow. Our results are consistent with detailed binary evolution models, and often lead to a smaller initial parameter space for stable mass transfer than other recipes in the literature. We use this method to investigate the origin of the Wolf-Rayet stars with O~star companions in the Small Magellanic Cloud, and find that the efficiency of the mass transfer process which lead to the formation of the Wolf-Rayet star was likely below 50\%.}


\maketitle

\section{Introduction}

Binary stars play a key role in stellar physics, since most massive stars, which enrich the universe with elements and shape star forming galaxies by their energy output, are part of a binary system \citep{1998NewA....3..443V,2012Sci...337..444S, 2017ApJS..230...15M}, which may lead to powerful phenomena like X-ray pulsars \citep{2006csxs.book..623T} and gravitational wave events \citep{2023pbse.book.....T}. Interacting binaries are an important site of stellar nucleosynthesis \citep{2009A&A...507L...1D,2021ARA&A..59..155M} and those of low mass are possible progenitors of Type~Ia supernovae, which are essential for mapping the Universe \citep{1998AJ....116.1009R,1998ApJ...507...46S,1999ApJ...517..565P}. 

Close binaries will sooner or later interact \citep{2012Sci...337..444S}, but what the outcome is when one component fills its Roche lobe is still under debate \citep{2012ARA&A..50..107L,2023arXiv231101865M}. A Roche-lobe overflow (RLO) where material is transferred steadily from one star to the other \citep[e.g.][]{1967ZA.....65..251K} is believed to lead to a stripped star in form of a hot sub-dwarf \citep{2020A&A...639L...6S,2020A&A...641A..43B} or a Wolf-Rayet (WR) star \citep{1989A&A...210...93L,2001A&A...369..939W} and a mass gainer which was spun up to fast rotation if the tides are negligible \citep{2013ApJ...764..166D,2020ApJ...888L..12W}, observable as a Be star \citep{2013A&ARv..21...69R}. If the RLO is dynamical unstable or the two components evolve into contact, a common envelope is formed which may be ejected \citep{2016A&A...596A..58K} or drives the system towards a stellar merger \citep{2013A&ARv..21...59I,2019Natur.574..211S}. Even if such a scenario is avoided, it is unclear how much of the material lost by the donor star is accreted by its companion \citep{2007A&A...467.1181D} and how much angular momentum the ejected material drains from the system \citep{1997A&A...327..620S}.

Several studies in the literature address these questions. A classical approach is to determine the mass-radius exponent of stellar models, which was done by e.g. \citet{1987ApJ...318..794H} for polytropes, and for more realistic stellar models by \citet{1989PhDT.........7H,1989SSRv...50..155H}  and most recently by \citet{2010ApJ...717..724G,2015ApJ...812...40G,2020ApJ...899..132G}. \citet{2001A&A...369..939W} examine the formation of contact in conservative detailed binary evolution models. \citet{2020A&A...638A..39L,2020ApJ...888L..12W,2022A&A...659A..98S} determine the accretion efficiency by letting the accretor take on mass until it reaches critical rotation and use an energy criterion to determine the outcome of a RLO.

\citet{kmh} and \citet{neo} showed that the accretor expands if the inflow of matter is too fast \citep[see also][]{1976ApJ...206..509U,1977ApJ...212..533F,1979A&A....75..255P,1989ApJ...341..306F}. If this expansion is large enough that the accretor also fills its Roche-lobe, one expects the formation of a common envelope, and after further expansion that material leaves the binary at the second Lagrange point (L2 overflow), which is expected to lead to the merger of the two stars as the expelled material drains a large amount of angular momentum from the system \citep{1976PASJ...28..593N}. \citet{1991A&A...241..419P} inferred from the work of \citet{kmh} that the accretor does not expand significantly as long as its accretion rate remains smaller than ten times its thermal timescale accretion rate. This approach is commonly used in many binary population studies \citep[e.g.][]{hurley02,2014ApJ...796...37S,2016ApJ...833..108S,2018MNRAS.481.4009V,2022ApJ...931...17V}, where the mass gain of the accretor is often limited by the thermal timescale accretion rate when the mass transfer rate exceeds this value (see \citet{1996A&A...309..179P} and \citet{2012A&A...546A..70T} for a more detailed approach). Other works do not consider the reaction of the accretor, and adopt more ad hoc merger criteria \citep[e.g.][]{2002ApJ...572..407B,2008ApJS..174..223B,2018MNRAS.481.1908K}.

In the last decades powerful codes like the one of \citet{1971MNRAS.151..351E,1972MNRAS.156..361E,1973MNRAS.163..279E,1973A&A....23..325E}, BEC \citep{2000ApJ...528..368H,2006A&A...460..199Y,2011A&A...530A.115B}, the Brussels codes STAREVOL and BINSTAR \citep{2000A&A...358..593S,2006A&A...448..717S,2006A&A...453..261P,2008A&A...489..395S,2013A&A...556A...4D,2013A&A...557A..40D,2013A&A...550A.100S}, and MESA \citep{2011ApJS..192....3P,2013ApJS..208....4P,2015ApJS..220...15P,2018ApJS..234...34P,2019ApJS..243...10P} were developed for the detailed modelling of both single and binary stars. However, even today it is still a large effort to model a complete stellar population with these one-dimensional approaches. Therefore rapid population synthesis codes have been developed, which approximate the evolution of binary systems either by analytic fits \citep[e.g.][]{hurley02} or by interpolation of precalculated single star models \citep[e.g.][]{2018MNRAS.481.1908K}. Thus, it is one aim of this study to provide a practical and efficient method to determine the outcome of a RLO without the need for detailed modelling.

Our study consists of two parts. In Sect.~\ref{sec1}, we analyse a generic set of accreting detailed single star models to find an accurate and practical description of their radius evolution, as function of the relevant parameters. The second part (Sect.~\ref{sec2}) is dedicated to the application of our results to grids of binary models, which we also compare to observation. Throughout the paper, we will compare our simplified models with detailed binary evolution models.

%
%

\section{Accretion on main-sequence stars}\label{sec1}

\subsection{Method}\label{sec1:method}

We calculated detailed single-star models of Small Magellanic Cloud (SMC) metallicity with initial chemical compositions as in \citet{2011A&A...530A.115B} and custom-built OPAL opacities \citep{1996ApJ...464..943I} in line with these abundances. The initial masses $M_\mathrm{i}$ of the models are 1, 1.5, 2, 3, 5, 7, 10, 15, 20, 30, 50, 70, and $100\msol$, and we assumed various constant accretion rates $\dot M$ (see Fig.~\ref{fig:hrd5} top, \ref{fig:hrd1}, and~\ref{fig:hrd2}), using MESA version 10108 \citep{2011ApJS..192....3P, 2013ApJS..208....4P, 2015ApJS..220...15P, 2018ApJS..234...34P}. We applied the Ledoux criterion for convection and used standard mixing-length theory with $\alpha_\mathrm{ml}=1.5$. We followed \citet{2019A&A...625A.132S} and \citet{2021A&A...653A.144H} and used semiconvection with $\alpha_\mathrm{sc}=10$ and a mass-dependent step-overshooting. We assume thermohaline mixing following \citet{2010A&A...521A...9C} with $\alpha_\mathrm{th}=1$. We simplify our models by treating them as non-rotating. After the stellar models have relaxed onto the main-sequence, they are subjected to a constant accretion of material that carries the same entropy as the model's uppermost mass shell. We let the models accrete until they have quintupled in mass, at which point we terminate the calculations.

\subsection{Results}\label{sec1:results}

\subsubsection{Behaviour of accreting main-sequence star models}\label{sec1:detmod}

\begin{figure}
    \includegraphics[width=\hsize]{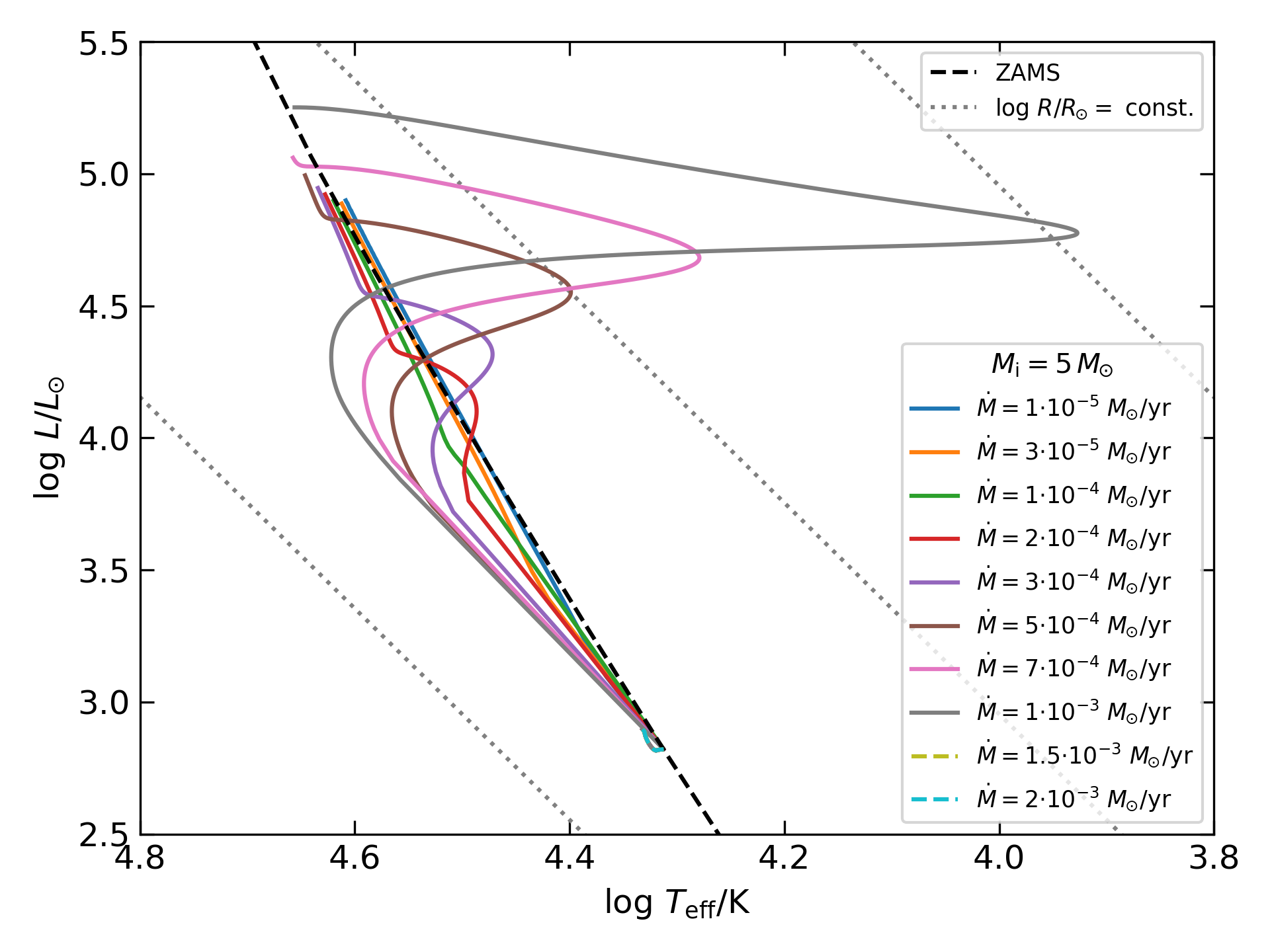}
    \includegraphics[width=\hsize]{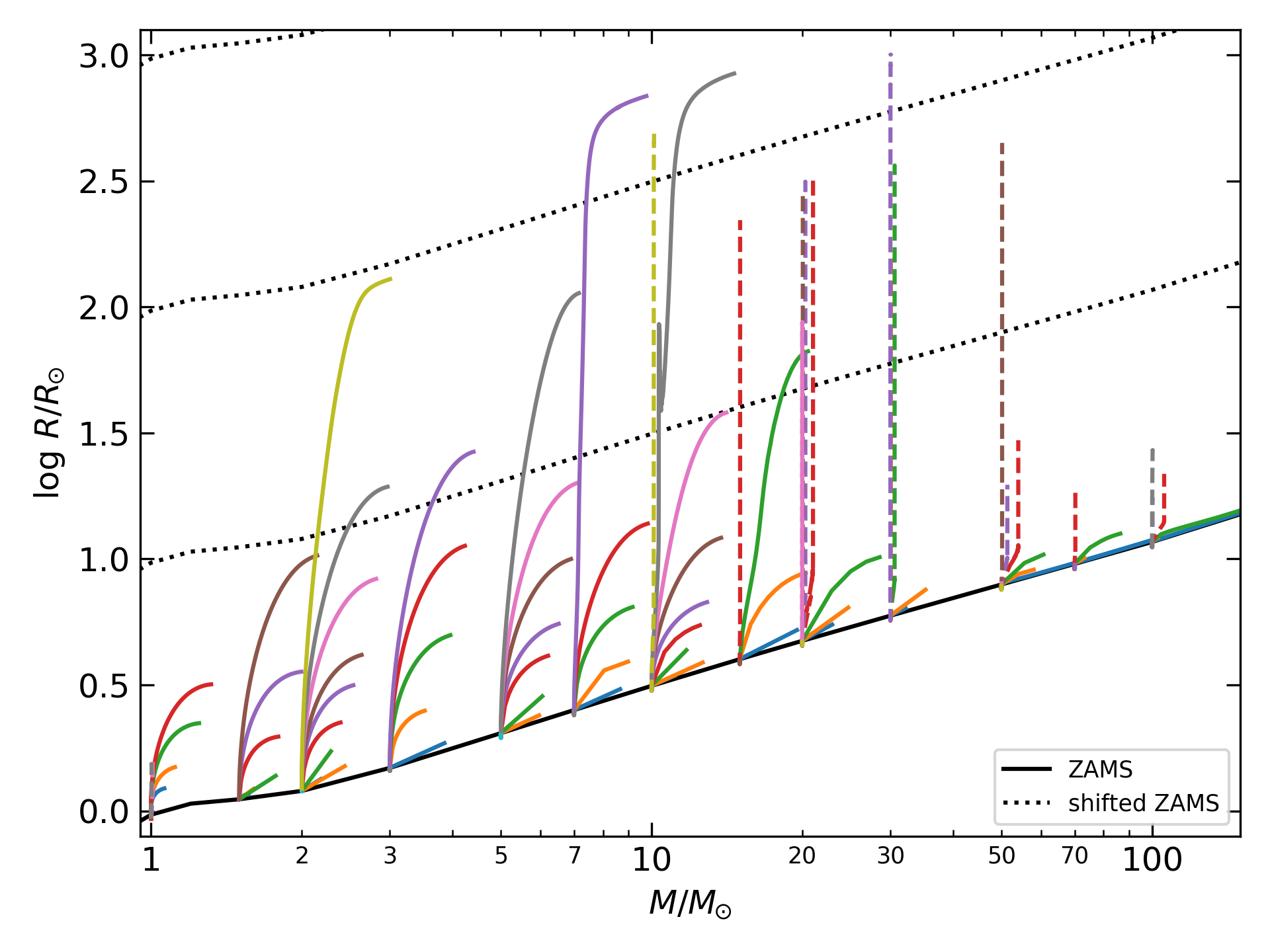}
    \caption{Evolution of our accreting single star models.
    Top: Evolution of the $5\msol$-models in the HRD for different accretion rates (indicated by colour). Stably swelling models are shown with solid lines and unstable models with dashed lines. We also show the ZAMS (black) and various lines of constant radius (grey).
    Bottom: Radius of accreting models as a function of mass. For each initial mass the colours are the same as in the corresponding HRDs (top panel, Fig.~\ref{fig:hrd1} and \ref{fig:hrd2}). Models that become unstable are indicated by dashed lines. Some are so short that they are hardly visible. The ZAMS radius is shown in black.}
    \label{fig:hrd5}
\end{figure}

We show the tracks of our $5\msol$-models in the Hertzsprung--Russell diagram (HRD) in Fig.~\ref{fig:hrd5} (top), which are typical for our model grid. We have indicated the different adopted accretion rates by colour. The corresponding diagram for the other initial masses are given in Fig.~\ref{fig:hrd1} and~\ref{fig:hrd2}. In the bottom panel we show the radius evolution as a function of the models' mass $M$ up to its maximum radius.

The stellar models with the lower accretion rates follow a common pattern. At the onset of accretion they briefly evolve to the left of the zero-age main-sequence (ZAMS) as a hydrodynamic response to the accretion only to rapidly increase their radius $R$ thereafter. The increase in radius is larger for higher accretion rates. After reaching a maximum radius $R_\mathrm{max}$ at a mass $M_{R=R_\mathrm{max}}$ the models contract again towards the ZAMS. From there they evolve along the ZAMS as they still accrete material. We call these models stable models or models with stable swelling.

The two models with the highest accretion rates of Fig.~\ref{fig:hrd5} (top, yellow and cyan lines) show a different evolution from the stable models, as their lines so short that they are barely visible in the two plots. In fact, these models barely accrete any mass and terminate shortly after the onset of the accretion due to numerical problems, as the time step becomes unreasonably small. We call these models unstable models or models with unstable swelling.

We observe the same patterns for the models of other masses (Fig.~\ref{fig:hrd1} and~\ref{fig:hrd2}). The larger the accretion rate, the larger is the maximum radius of the models, and beyond a certain accretion rate the models terminate due to numerical problems. For initial masses above $10\msol$, the unstable models show a large increase in radius before they terminate (Fig.~\ref{fig:hrd5}, bottom, dashed lines). This extends the interpretation of the unstable models to the description that, above a certain accretion rate, a small increase in mass causes the star to expand by a very large amount. Consider, for instance, the models with an initial mass of $15\msol$ in Fig.~\ref{fig:hrd5} (bottom). From $1\cdot10^{-3}\mpa$ (blue) to $3\cdot10^{-3}\mpa$ (green) the swelling becomes stronger, but at an accretion rate of $6\cdot10^{-3}\mpa$ (red) the mass-radius-curve becomes nearly vertical. Furthermore, for initial masses above $10\msol$ we find that with increasing mass a stable radius increase becomes less possible, and from $30\msol$ on the models either accrete stably along the ZAMS or become unstable.

Between the stable and unstable regimes we find three borderline models: $2\msol$ with $\dot M = 2\cdot10^{-4}\mpa$, $7\msol$ with $\dot M = 2\cdot10^{-3}\mpa$, and $10\msol$ with $\dot M = 2.5\cdot10^{-3}\mpa$. These models have in common (in contrast to the other stable ones) that they reach their maximum radius as red (super) giants, and that their tracks in the HRD depict a lower curvature at maximum radius. In Fig.~\ref{fig:hrd5} (bottom) we see that the three boundary models reach the largest radii of all stable models of the same initial mass, and show a plateau in this. We assume that the unstable models would show the same behaviour as the borderline models if they could be calculated further, which is supported by the binary models presented in Sect.~\ref{sec1:test}.

\begin{figure}
    \includegraphics[width=\hsize]{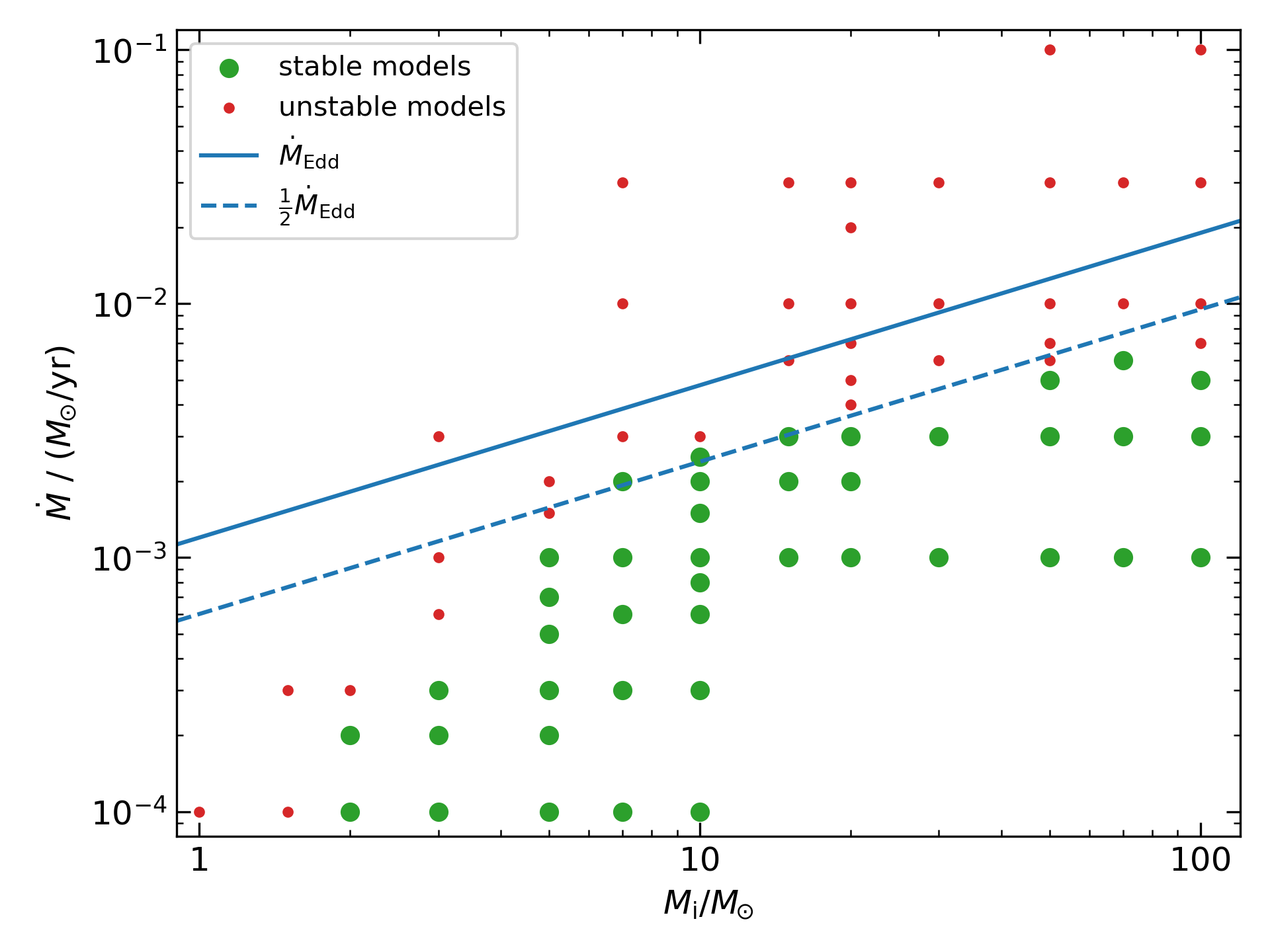}
    \caption{Initial mass and adopted accretion rate for our stable (green) and unstable (red) models. Not all models are shown as we focus on the borderline between stable and unstable models at high masses. The blue lines indicate the Eddington accretion rate assuming electron-scattering opacity.}
    \label{fig:edd}
\end{figure}

We can understand whether a model accretes stably or not by considering the Eddington accretion rate, given by 
\begin{equation}
    \dot M_\mathrm{Edd} = \frac{4\pi cR}{\kappa},
\end{equation}
where $c$ is the speed of light, $R$ is the stellar radius and $\kappa$ the opacity \citep{2023pbse.book.....T}. By assuming $R \propto M^{0.6}$, which fits our ZAMS models well \citep{2013sse..book.....K} and $\kappa=0.34\,\mathrm{cm^2/g}$, which is a good approximation for $M>10\msol$, we find 
\begin{equation}
    \dot M_\mathrm{Edd} \approx \left[1.2\cdot10^{-3} \mpa\right] \cdot \left(\frac{M}{\msol}\right)^{0.6}
\end{equation}
In Fig.~\ref{fig:edd} we show the initial mass and the adopted accretion rate of our models. The blue continuous line shows the Eddington accretion rate and the dashed line assumes an opacity of twice the electron-scattering value, which approximates the total opacity in the outer envelope of our models. The latter matches well to the borderline between the stable and the unstable models. This means, that our models become unstable if the accretion rate exceeds the Eddington accretion rate of the accretor.

\subsubsection{Analytic fits}\label{sec1:fit}

\begin{figure}
    \includegraphics[width=\hsize]{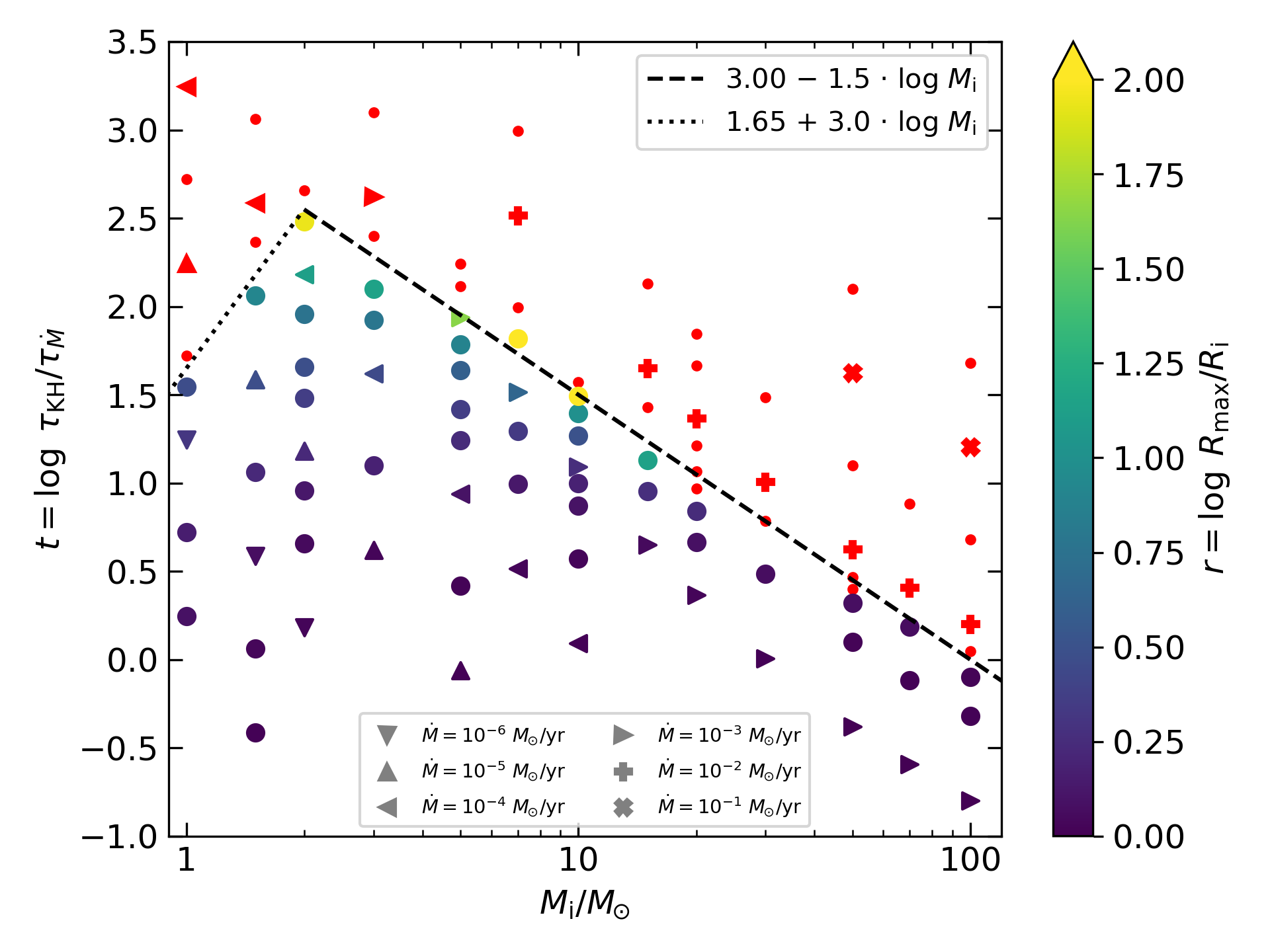}
    \caption{Maximum radius of accreting models as a function of initial mass and the ratio of thermal and mass transfer timescales. Unstable models are shown in red. Special symbols are used for selected accretion rates. The two black lines distinguish between stable and unstable models.}
    \label{fig:Rmax}
\end{figure}

To provide a description of the response of the stellar models to accretion, Fig.~\ref{fig:Rmax} shows a synopsis of our models. We express the accretion rate in terms of the logarithmic ratio 
\begin{equation}\label{eq:t}
    t = \log\frac{\tau_\mathrm{KH}}{\tau_{\dot M}}
\end{equation}
of the thermal timescale $\tau_\mathrm{KH}$ and the mass gain timescale $\tau_{\dot M}$ at the onset of accretion, which are defined as
\begin{align}
    \tau_\mathrm{KH} &= \frac{3GM^2}{4RL}, \label{eq:kh}\\
    \tau_{\dot M} &= \frac{M}{\dot M}, \label{eq:md}
\end{align}
where $L$ is the luminosity and $G$ is the gravitational constant \citep{2004sipp.book.....H,2013sse..book.....K}. We find that the boundary between stable and unstable models can be described by two simple lines, which are given by 
\begin{equation}\label{eq:grenze}
    t_\mathrm{max} = \begin{cases} 1.65 + 3\log \Mzams, & \Mzams < 2\msol \\ 3-1.5 \log \Mzams, & \Mzams > 2\msol.\end{cases}
\end{equation}
For masses above $2\msol$ we find that our models become more unstable with increasing initial mass. Models below that show the opposite trend. Also, the closer a model of a given initial mass is to the boundary, the larger is its maximum radius. For a fixed $t$, the maximum radius is minimal around $\Mzams=2\msol$. Near the boundary, the maximum radii are up to 100 times larger than the ZAMS radii. For models with $t = \ltt < 0$ the swelling is negligible.

\begin{figure}
    \includegraphics[width=\hsize]{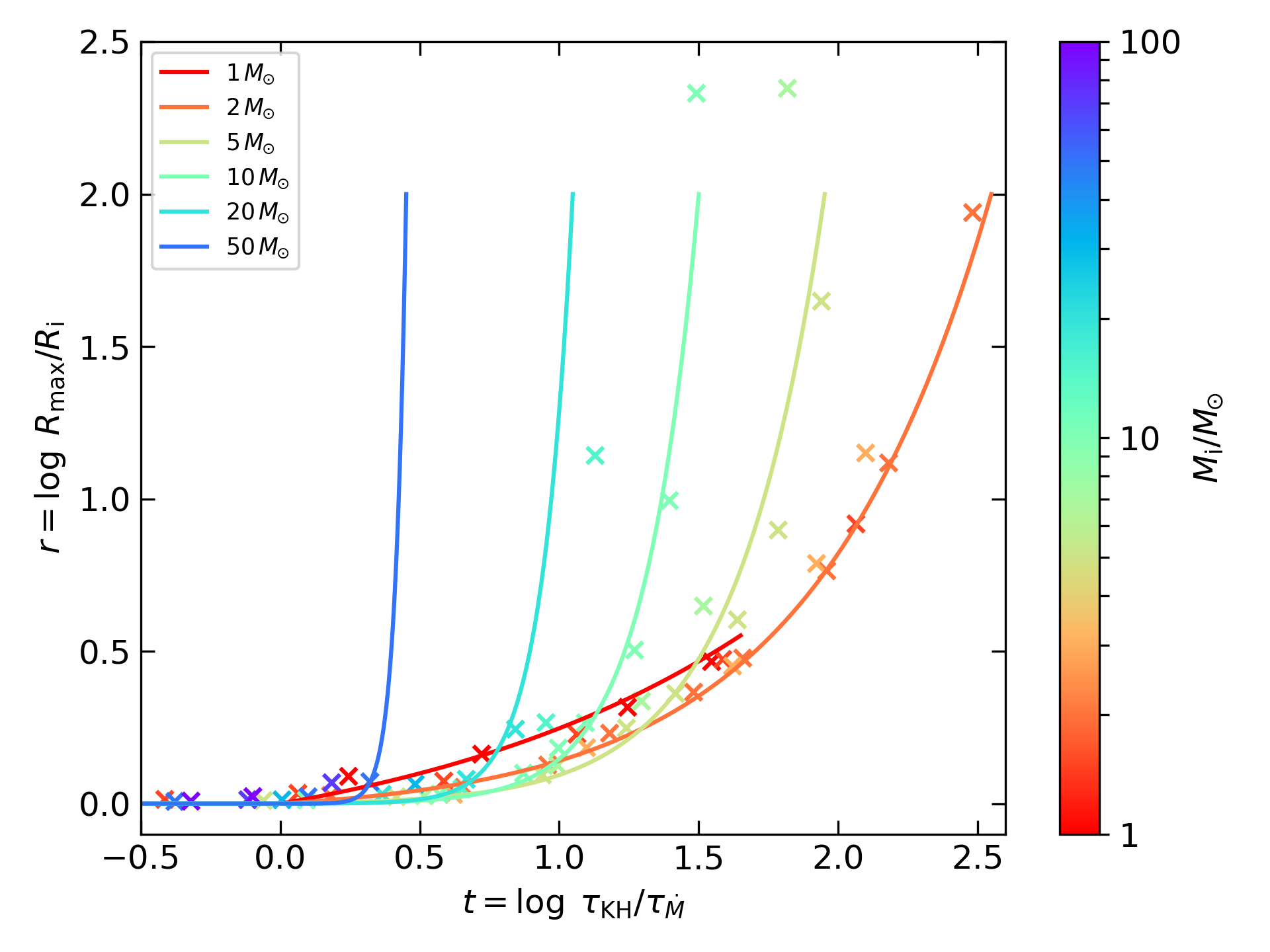}
    \includegraphics[width=\hsize]{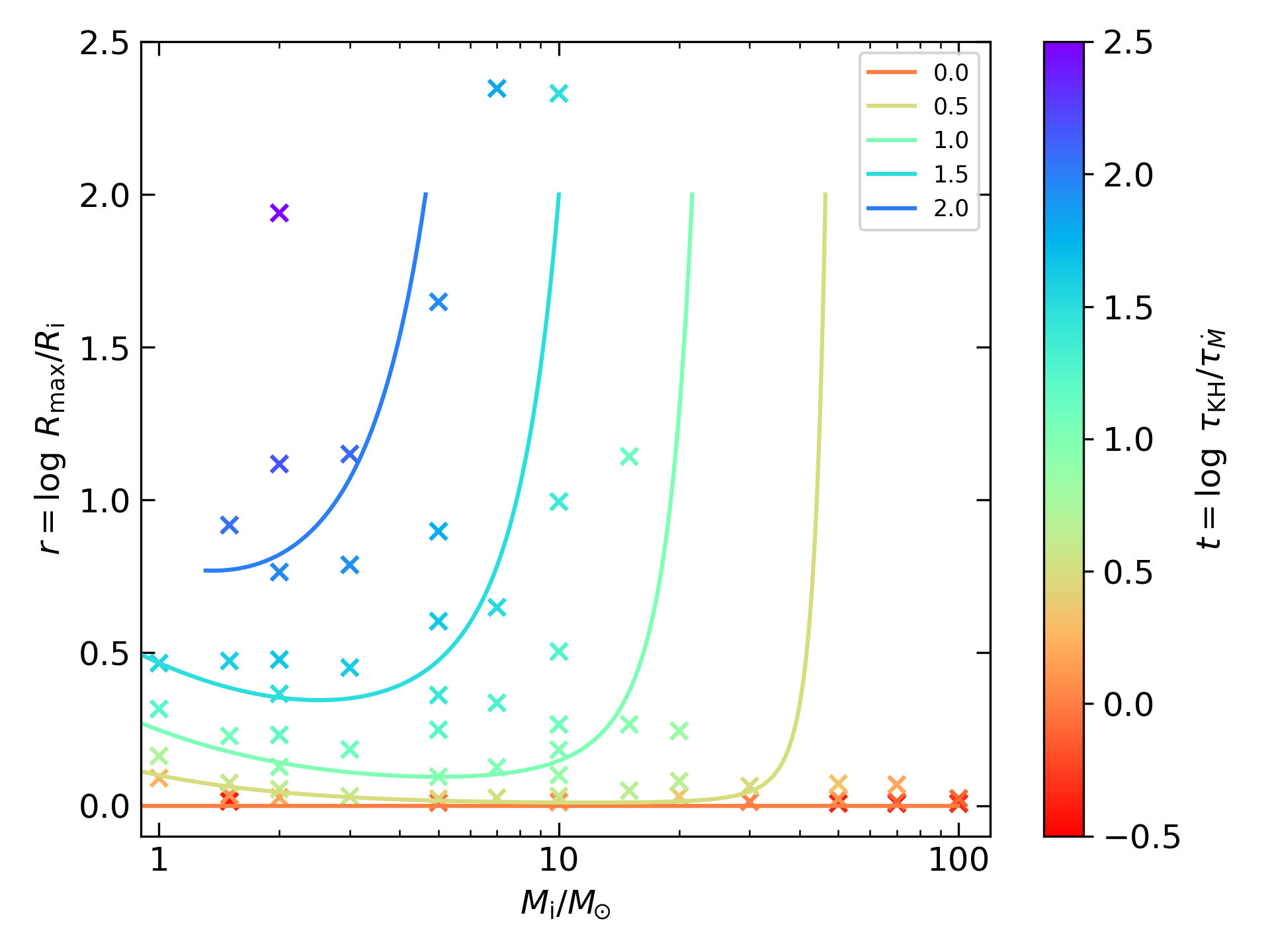}
    \caption{Logarithmic radius increase $r$ of accreting models as a function of initial mass and the ratio of thermal and mass transfer timescale together with our fit (lines) for selected initial masses (top) and for selected timescale ratios (bottom).}
    \label{fig:Rfit}
\end{figure}

\begin{figure}
    \includegraphics[width=\hsize]{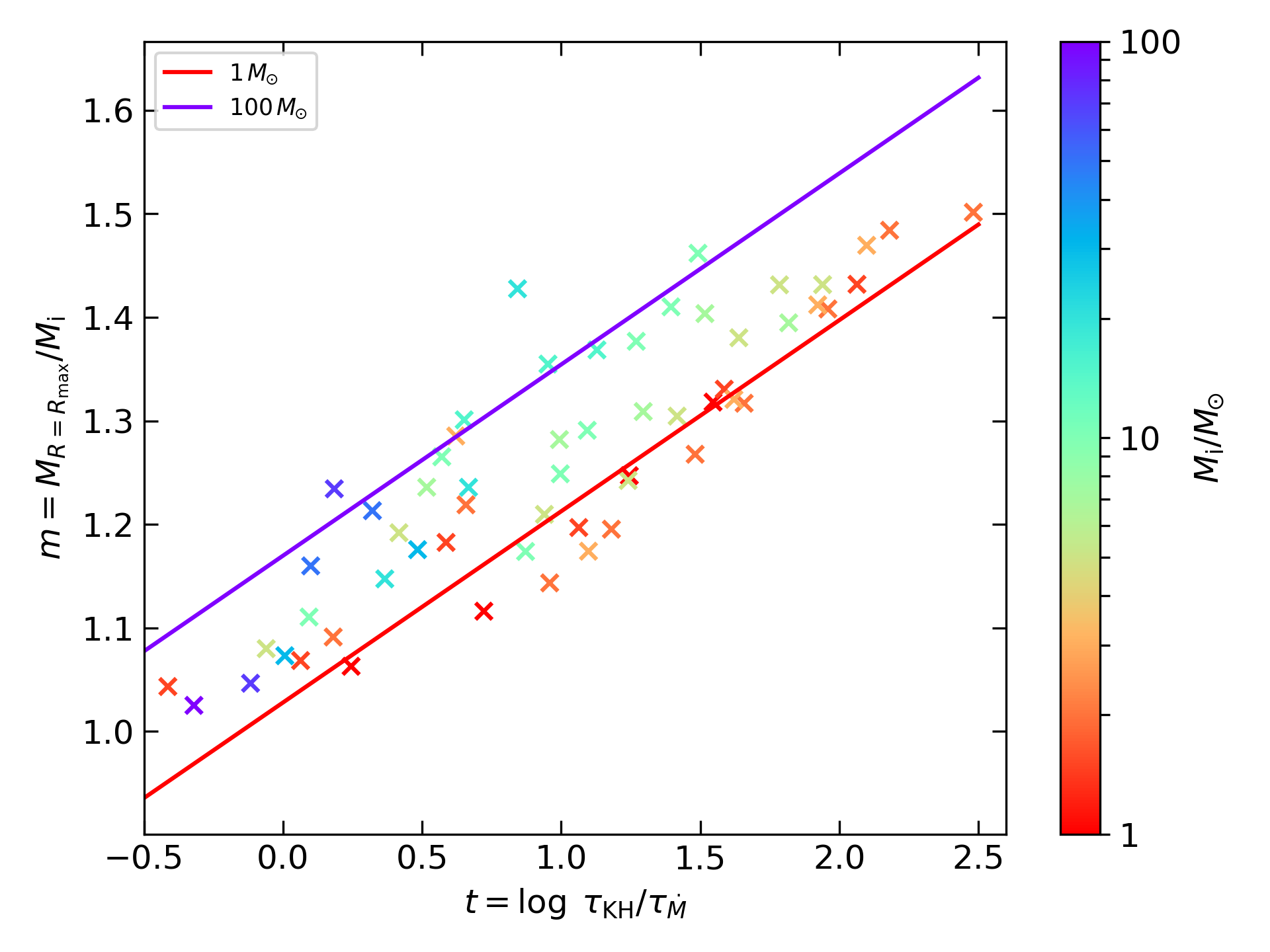}
    \includegraphics[width=\hsize]{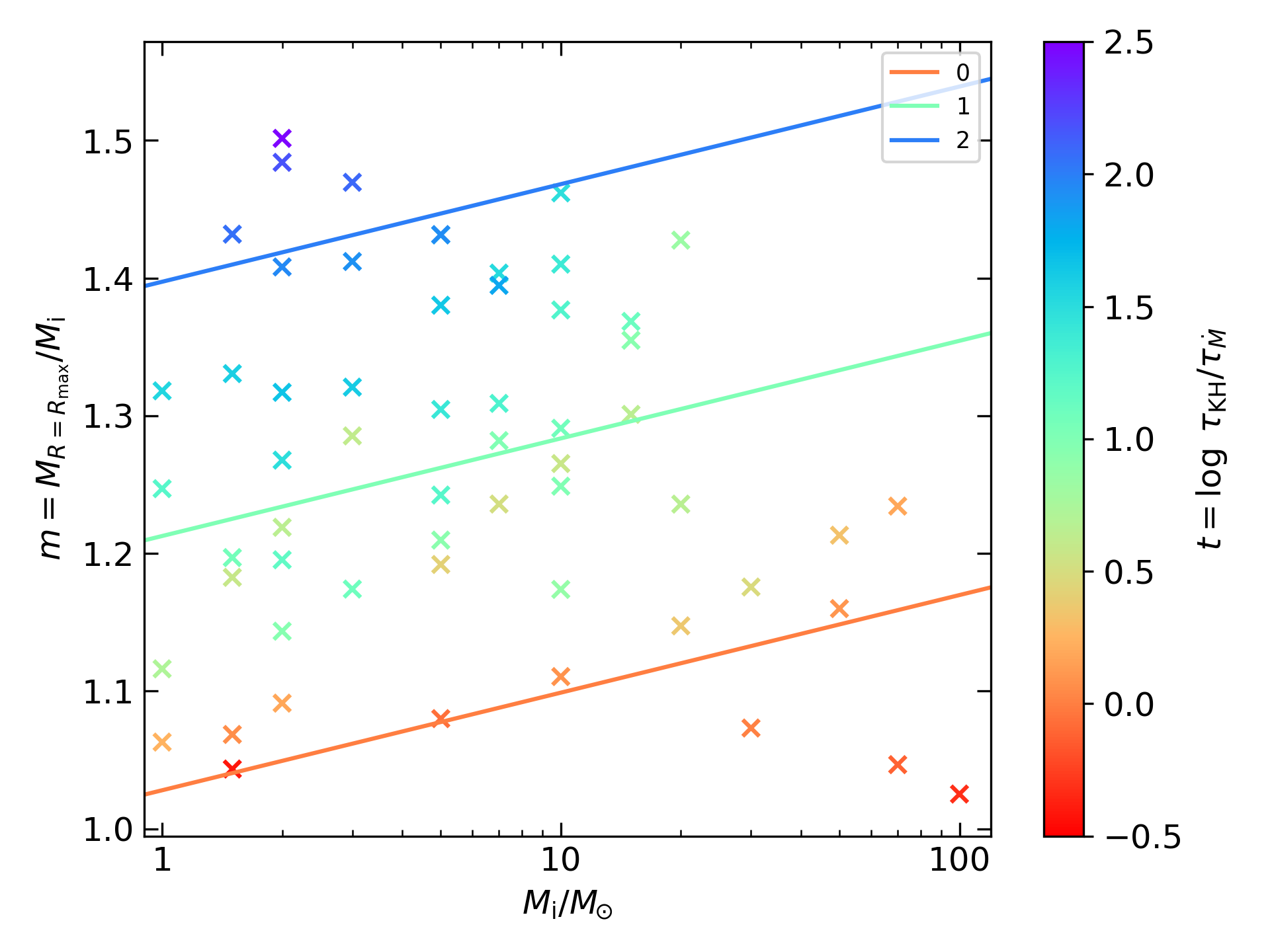}
    \caption{Mass at which accreting models reach their largest radius as a function of initial mass and the ratio of thermal and mass-transfer timescale together with our fit for selected initial masses (top) and for selected timescale ratios (bottom).}
    \label{fig:Mfit}
\end{figure}

In order to incorporate our results into a rapid binary population synthesis code, we fitted simple functions to the logarithmic radius increase $r = \log R_\mathrm{max}/R_\mathrm{i}$ and the mass increase to reach the maximum radius $m = M_{R=R_\mathrm{max}}/\Mzams$ as functions of the initial mass $\Mzams$ and the logarithmic timescale ratio $t$. For the function $r$ we require that $r(t=0)=0$ and that it reaches values of 2 at the boundary towards the unstable models, which roughly corresponds to the radius increase of the three borderline models.

\begin{table}
    \centering
    \caption{Parameters found for Eq.~\ref{eq:Rfit} and~\ref{eq:Mfit} as well as the root-mean-square relative deviation $\delta_\mathrm{rms}$ and the maximum relative deviation $\delta_\mathrm{max}$.}
    \begin{tabular}{ccc} \hline\hline
        & Eq.~\ref{eq:Rfit} & Eq.~\ref{eq:Mfit} \\ \hline
        $a$ & $5.44\pm0.53$ & $0.0709\pm0.0149$ \\
        $b$ & $2.38\pm0.27$ & $0.185\pm0.011$ \\
        $c$ & -- & $1.028\pm0.020$ \\ \hline
        $\delta_\mathrm{rms}$ & 0.28 & 0.04 \\
        $\delta_\mathrm{max}$ & 1.60 & 0.11 \\ \hline
    \end{tabular}
    \label{tab:fit}
\end{table}

We find a good fit for $R_\mathrm{max}$ with the function
\begin{equation}\label{eq:Rfit}
    r(\log \Mzams, t) = 2 \cdot \frac{\exp{\left[(a\log\Mzams+b) \frac{t}{3-1.5\log\Mzams} \right]}-1}{\exp{\left[a\log\Mzams+b\right]}-1}
\end{equation}
with the parameters $a$ and $b$ given in Table~\ref{tab:fit}. Note that $t$ is not divided by $t_\mathrm{max}$ as in Eq.~\ref{eq:grenze}, but only by the second line of the formula. We show the values of the detailed models as well as our fit function in Fig.~\ref{fig:Rfit}. The data and the fit are in good agreement, as the root-mean-square relative deviation and the maximum relative deviation are reasonably small (Table~\ref{tab:fit}).

For the mass at the maximum radius, we find the linear function given by
\begin{equation}\label{eq:Mfit}
    m (\log \Mzams, t) = a \log \Mzams + b t + c
\end{equation}
fits well. The parameters $a,b,c$ are listed in Table~\ref{tab:fit}. The values of the detailed models and the fit are shown in Fig.~\ref{fig:Mfit}.

\subsection{Discussion}\label{sec1:discuss}
In this section we discuss the uncertainties of our prescription (Sect.~\ref{sec1:uncert}) and compare it with detailed binary models (Sect.~\ref{sec1:test}) and previous works (Sect.~\ref{sec1:prev}).

\subsubsection{Uncertainties}\label{sec1:uncert}

In our models we have omitted stellar rotation for simplicity. However, rotation is a ubiquitous phenomenon observed in stars \citep{2000ARA&A..38..143M,2012ARA&A..50..107L}. It has several effects on stars. First of all, there is the deformation of the stellar surface due to the centrifugal force \citep[e.g.][]{2013sse..book.....K}. This effect can increase the equatorial radius by a factor of up to 1.5. This number is within the uncertainty of our prediction for the accretion induced swelling (Sect.~\ref{sec1:results}). Moderate rotation does not alter stellar evolution much \citep{2011A&A...530A.115B,2016ApJ...823..102C}, so the expected corotation in close binaries \citep{2009A&A...497..243D} and the fast rotating branch of the main sequence \citep{2013A&A...550A.109D,2020ApJ...888L..12W} are not affected.

On the other hand it is generally accepted that mass transfer leads to a spin-up of the accretor star to close to critical rotation \citep{2013ApJ...764..166D}. \citet{1981A&A...102...17P} showed, that only a small amount of matter is required to spin up the star to critical rotation. Thus all our models should rotate critically very soon. The induced rotational mixing \citep{2000ApJ...528..368H,2005ApJ...626..350H} may help to incorporate the incoming material into the star and thus prevent or enhance the swelling of the accretor. The first case this is supported by the observation that an accreting stellar model close to the Hayashi lines (where models are fully convective and thus well mixed) can accrete a large amount of matter with only a small further swelling (e.g. Fig~\ref{fig:hrd5}, bottom, purple line starting at $7\msol$). Stellar rotation might also affect the orbital evolution through spin-orbit coupling by tides \citep[e.g.][]{1977A&A....57..383Z}.

It is likely that the accretion rate is not constant, but varies as in the models of, \citet{2020A&A...638A..39L}, \citet{2020ApJ...888L..12W} and \citet{2022A&A...659A..98S}, in their case due to the rotational inhibition of accretion. The variation can also be due to the evolution of the Roche lobe radius. For a strong variation, as in the former example, probably only the time interval with the most efficient accretion matters, and \citet{2015ApJ...812...40G} argue that the latter effect is negligible. Also, for a strongly varying accretion rate, the assumption we will make in Eq.~\ref{eq:tapprox} is no longer valid.

We have assumed that the material arriving at the accretor has the same entropy as its surface. \citet{2015ApJS..220...15P} give a timescale argumentwhich indicates that this is always true. However, according to the stellar structure equations \citep[e.g.][eqs.~(4.45) to~(4.47)]{2013sse..book.....K} a change in entropy affects the luminosity and, by the Stefan-Boltzmann law, the radius of the star. Whether the swelling is enhanced or diminished depends on the actual entropy of the material, which is controlled by the initial entropy and the accretion mechanism, and requires further analysis. Whether the swelling of the star can lead to a change from accretion via a Keplerian disk to direct impact \citep{2005A&A...435.1013P} and the consequences to the arriving entropy need to be addressed.


\subsubsection{Comparison with detailed binary models}\label{sec1:test}

\begin{figure}
    \centering
    \includegraphics[width=\hsize]{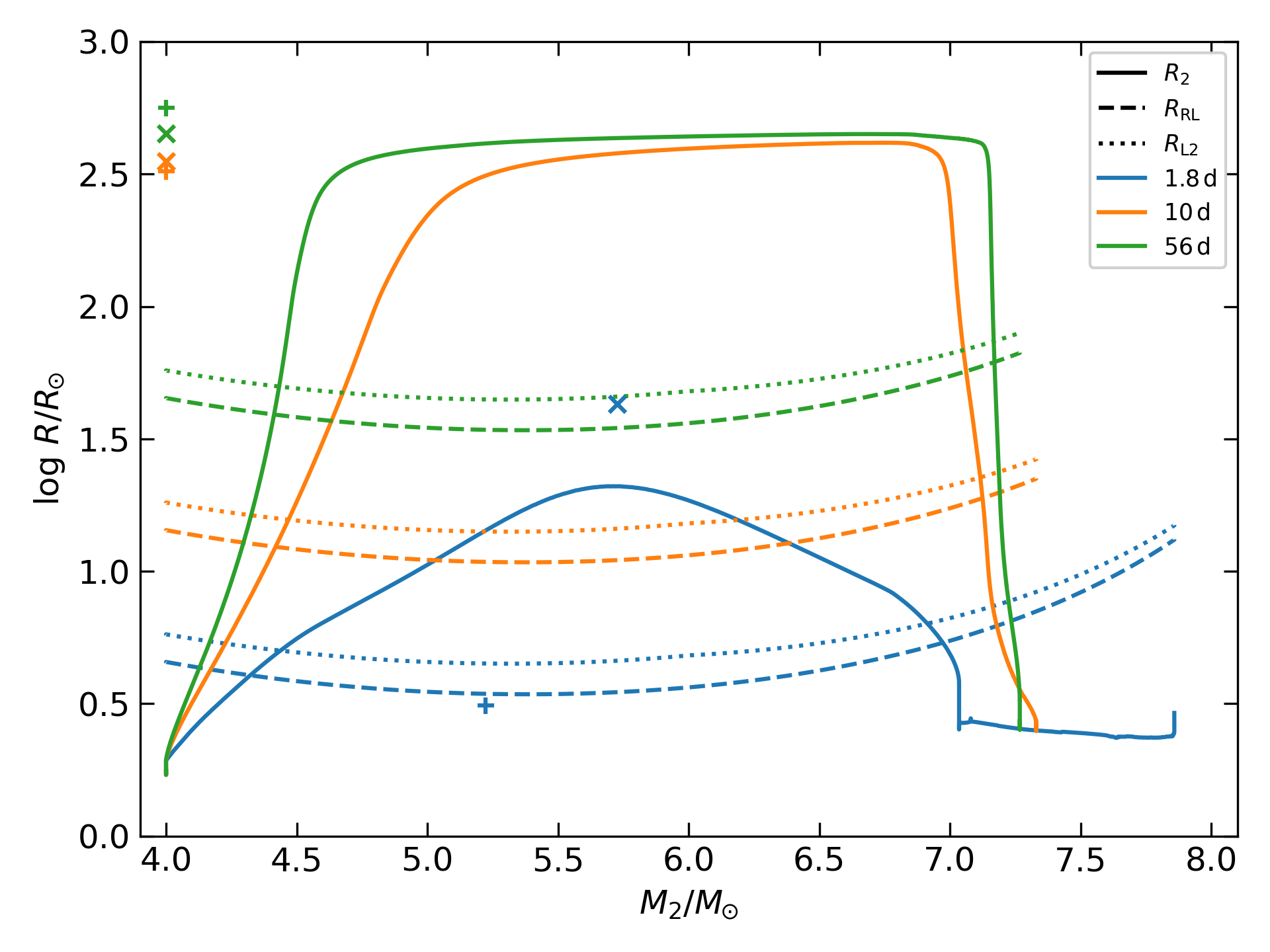}
    \includegraphics[width=\hsize]{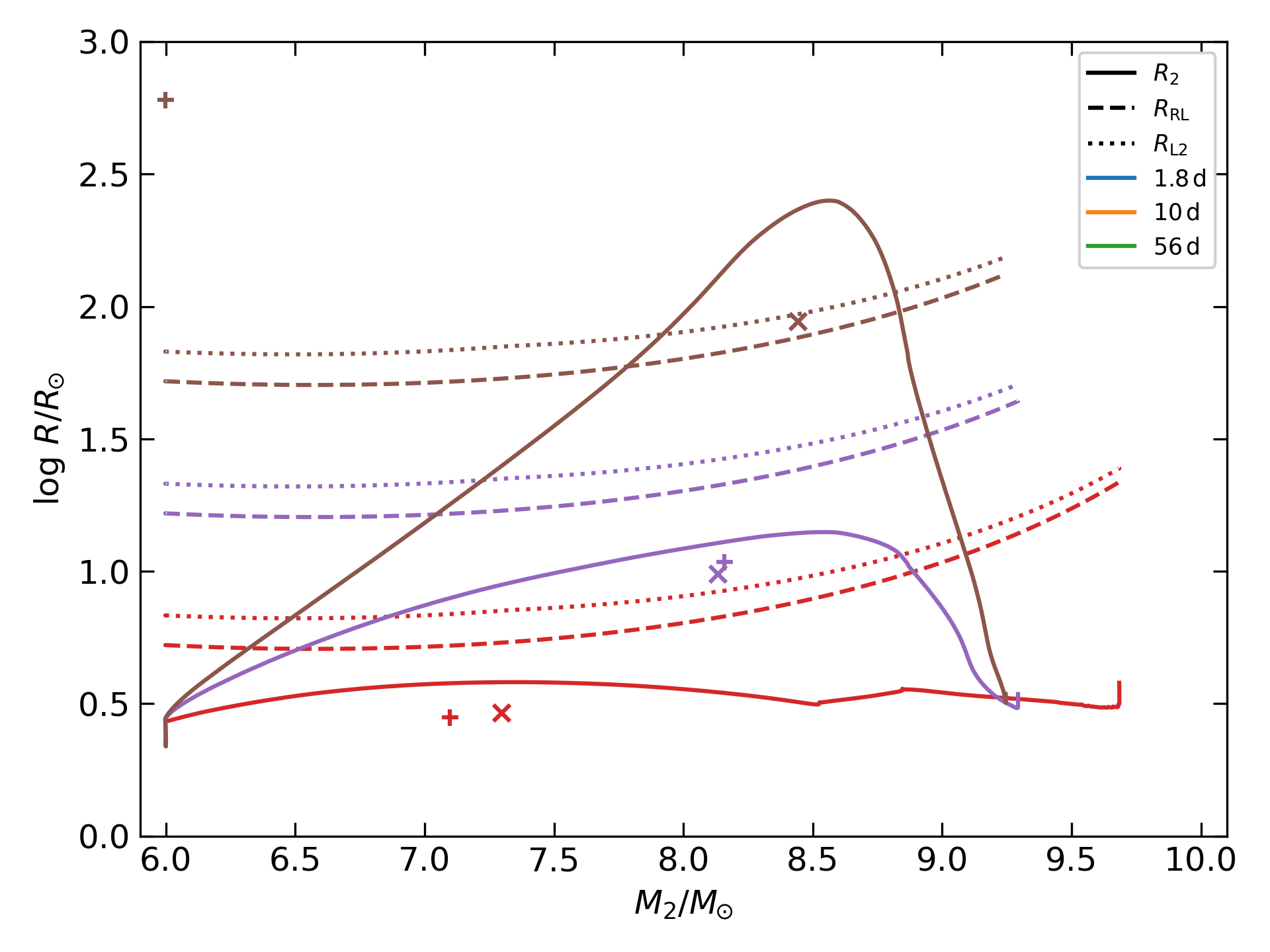}
    \includegraphics[width=\hsize]{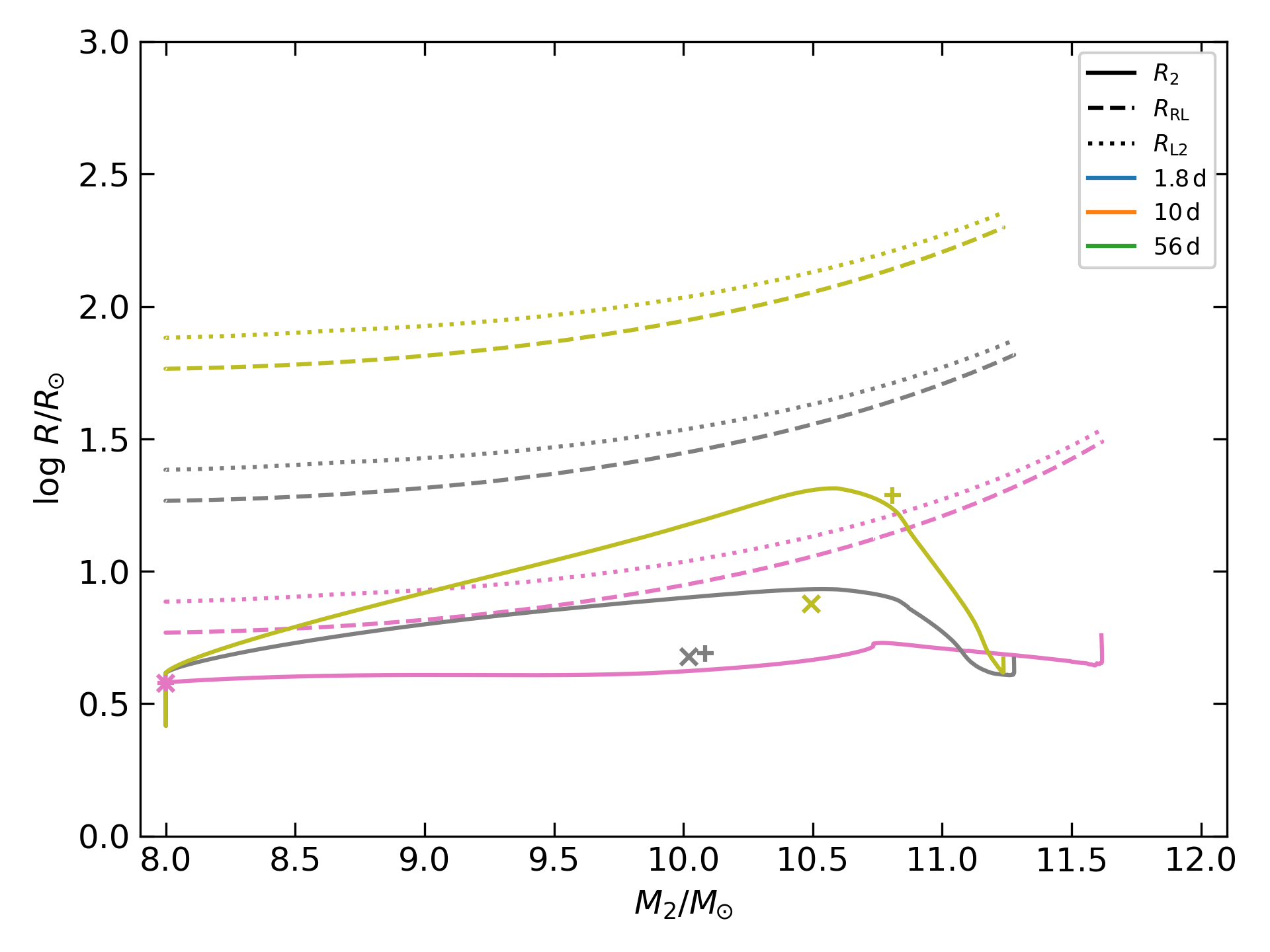}
    \caption{Mass-radius evolution (solid line) of detailed accretor models for different initial periods (colour) and initial accretor masses (top $4\msol$, middle $6\msol$, bottom $8\msol$). The donor always has an initial mass of $10\msol$. The size of the accretor's Roche-lobe and L2-sphere are shown as dashed and dotted lines. We have used $\times$-symbols to indicate the maximum of the mass-radius curve $(R_\mathrm{max},M_{R=R_\mathrm{max}})$ according to Eq.~\ref{eq:Rfit} and~\ref{eq:Mfit}, based on the maximum mass transfer rate of the detailed model. The $+$-symbols indicate the same, but for an estimate of the mass transfer rate based on the conditions just before the RLO (Eq.~\ref{eq:tapprox}). If a symbol is placed at a high radius but at the initial mass of the model, the model is expected to swell unstably and the radius is the Hayashi radius (see Sect.~\ref{sec2:met}).}
    \label{fig:test}
\end{figure}

\begin{table*}
    \centering
    \caption{Initial masses $M_\mathrm{1i}$ and $M_\mathrm{2i}$ and initial orbital periods $P_\mathrm{i}$ of our detailed binary models and the case of RLO through which they pass. We distringuish between Case~B with a donor with radiative (Br) and convective (Bc) envelope. We display the outcomes of the RLO according to the detailed binary models (Sect.\ref{sec1:test}), the radius estimate using the maximum mass transfer rate of the detailed binary models ($\times$ in Fig.~\ref{fig:test}), the radius estimate based on Eq.~\ref{eq:tapprox} ($+$ in Fig.~\ref{fig:test}), and the outcome from Sect.~\ref{sec2:res}. We qualify the stability of the swelling and the occurrence of L2 overflow (L2O), contact, or neither (donor stripping).}
    \label{tab:bin}
    \begin{tabular}{cccccccccc} \hline\hline
        &&&&& outcome & outcome & outcome & outcome \\
        \# & $M_\mathrm{1i}/\msol$ & $M_\mathrm{2i}/\msol$ & $P_\mathrm{i}/\days$ & Case & det. model & with max$(\dot M)$ & with Eq.~\ref{eq:tapprox}  & of Sect.~\ref{sec2:res} \\ \hline
         1 & 10 &  4 & 1.8 & A  & stable, L2O & stable, L2O & stable, stripping & L2O \\
         2 & 10 &  4 &  10 & Br & unstable, L2O & unstable, L2O & unstable, L2O & L2O \\
         3 & 10 &  4 &  56 & Br & unstable, L2O & unstable, L2O & unstable, L2O & L2O \\
         4 & 10 &  4 & 316 & Bc & no solution & unstable, L2O & unstable, L2O & L2O \\
         5 & 10 &  6 & 1.8 & A  & stable, stripping & stable, stripping & stable, stripping & stripping\\
         6 & 10 &  6 &  10 & Br & stable, stripping & stable, stripping & stable, stripping & stripping\\
         7 & 10 &  6 &  56 & Br & stable, L2O & stable, contact & unstable, L2O & L2O\\
         8 & 10 &  6 & 316 & Bc & no solution & unstable, L2O & unstable, L2O & L2O\\
         9 & 10 &  8 & 1.8 & A  & stable, stripping & stable, stripping & stable, stripping & stripping\\
        10 & 10 &  8 &  10 & Br & stable, stripping & stable, stripping & stable, stripping & stripping\\
        11 & 10 &  8 &  56 & Br & stable, stripping & stable, stripping & stable, stripping & L2O\\
        12 & 10 &  8 & 316 & Bc & no solution & unstable, L2O & unstable, L2O & L2O\\
        13 & 30 & 12 & 1.8 & A  & unstable, L2O & stable, L2O & stable, stripping & stripping \\
        14 & 30 & 12 &  10 & Br & unstable, L2O & unstable, L2O & unstable, L2O & L2O \\
        15 & 30 & 12 &  56 & Br & unstable, L2O & unstable, L2O & unstable, L2O & L2O \\
        16 & 30 & 12 & 316 & Br & unstable, L2O & unstable, L2O & unstable, L2O & L2O \\ \hline
    \end{tabular}
\end{table*}

To validate our results, we compare our fits with detailed binary models undergoing RLO calculated with MESA (see Sect.~\ref{sec1:method}). The adopted initial masses and initial periods are listed in Table~\ref{tab:bin}. We use the same physical assumptions as in Sect.~\ref{sec1:method}, and the structure of both binary components is
calculated in parallel with the evolution of the orbit. The models are assumed to be non-rotating, as we are not interested in the radius increase due to this effect. We assumed a constant accretion efficiency of $\eps=50\%$, and that the ejected material carries the specific angular momentum of the accretor \citep{1997A&A...327..620S}. We have used the mass transfer scheme \texttt{roche\_lobe}. To be able to measure the maximum accretor radius, we allow the accretor to overfill its L2-volume without losing mass or terminating the calculation.

The mass-radius evolution of the $10\msol$ accretors is shown in Fig.~\ref{fig:test} together with estimates according to Eq.~\ref{eq:Rfit} and~\ref{eq:Mfit} based on the maximum mass transfer rate of the models ($\times$-symbols). In general we find satisfactory agreement. Typically we miss the maximum radius by no more than a factor of 2. Models \#2 (orange) and \#3 (green) stay at large and relatively constant radii for a while, similar to the borderline models mentioned in Sect.~\ref{sec1:detmod}. They also swell unstably according to Eq.~\ref{eq:Rfit} and~\ref{eq:Mfit}. Models \#4, \#8, \#12 are not shown because the calculations terminated due to numerical problems (time step limit) shortly after the onset of RLO. Typically, our recipe yields smaller accretor radii than in detailed calculations, making it a rather conservative estimate of L2 overflow.

There are two main reason for the differences between our approach and the detailed models. Our fitting function (Sect.~\ref{sec1:fit}) is very steep, and thus small uncertainties can lead to large changes in the resulting maximum radius. This could be improved with a denser model grid and a refined fit. Second, the accretion rate imposed by the donor is time dependent and a notable deposition of material on the mass gainer before the maximum mass transfer rate is reached could change our prediction. Considering a time dependent mass transfer rate is beyond the scope of our approach.

\subsubsection{Comparison with previous work}\label{sec1:prev}

Numerical experiments for accreting stars have been carried out by \citet{kmh} and \citet{neo}. They arrive at the same qualitative result as our study, namely that the maximum stellar radii increase as the accretion rate increases. If we compare the tracks in our HRDs with those of \citet[][their fig.~1-3]{kmh} and \citet[][fig.~1]{neo}, we find that the evolutionary tracks of \citet{neo}, like ours, intersect each other for a given initial mass, but those of \citet{kmh} do not. On the other hand, we find similarities between both fig.~4 of \citet{kmh} and fig.~4 of \citet{neo} and our Fig.~\ref{fig:hrd5} (bottom), not only in shape of the tracks but also in terms of critical accretion rate. \citet{neo} find that an accretion rate greater than $4\cdot10^{-3}\mpa$ is required for a $20\msol$ model to be unstable. We find a slightly lower rate of $3\cdot10^{-3}\mpa$. Similarly, for the $5\msol$ and the $10\msol$ models, we also find slightly lower critical accretion rates compared to \citet{kmh}. The differences could be caused by the used opacities, first because the old models did not include the iron-peak opacity and secondly we used a lower metallicity, which also enters the opacity and thus the Eddington limit.

\citet{1991A&A...241..419P,1994A&A...288..475P} state that the response of the accretor becomes important when the thermal timescale is ten times larger than the accretion timescale, i.e. for $t=1$. In contrast, we find that the radius of the accretor already deviates from equilibrium when the accretion timescale is equal to the thermal timescale ($t=0$) and the swelling becomes unstable at $t=t_\mathrm{max}$. Our $t_\mathrm{max}$ is mass dependent in contrast to their mass independent limit of $t=1$. However, these studies and subsequent work \citep[e.g.][]{2014ApJ...796...37S,2016ApJ...833..108S,2015ApJ...805...20S} assume that the unstable swelling leads to a reduced accretion efficiency rather than a merger.


Very recently, a similar study was put forward by \citet{2024arXiv240406148Z}, who calculated accreting models at Solar metallicity. Their models behave qualitatively as our models, but a close inspection reveals, that our models swell less for the same accretion rate. For example, our $5\msol$-model with $\dot M = 10^{-3}\mpa$ reaches about $120\rsol$, while theirs swells to around $600\rsol$. This behaviour may relate to the metallicity dependence of the opacity (cf. Sect.~\ref{sec1:detmod}). A higher metallicity increases the opacity, which in turn decreases the Eddington accretion rate. This means that for increasing metallicity the borderline between stable and unstable accretion moves to smaller timescale ratios $t$. The magnitude of the shift is hard to estimate from the combined model data. Their $5\msol$-model with $\dot M = 10^{-3}\mpa$ appears to be what we call a borderline model, our comparable model accretes stably and our $5\msol$-model with $\dot M = 1.5\cdot10^{-3}\mpa$ is unstable. Form that we can estimate that the shift of the borderline between stable and unstable accretion may be less than $0.2\,\mathrm{dex}$ when going from SMC to Solar metallicity. 

\section{Predictions for L2 overflow}\label{sec2}

In this Section, we apply our model for the swelling of the accretor star to model grids of binary systems. We follow the evolution of the accretor radius and of its Roche radius. We then determine the conditions under which the accretor overfills its Roche lobe, leading at first to contact and with further overfilling to an L2 overflow. We assume that the latter leads to a merger of the two stars \citep{1976PASJ...28..593N}.

\subsection{Method}\label{sec2:met}

We model the binary systems in our grid based on detailed non-rotating single star models computed with MESA version 10108 \citep{2011ApJS..192....3P, 2013ApJS..208....4P, 2015ApJS..220...15P, 2018ApJS..234...34P}. The initial masses are 8, 10, 15, 20, 30, 50, 70, and $100\msol$, and the physical assumptions are identical to those in Sect.~\ref{sec1:method}, unless otherwise stated. We use overshooting ($\alpha_\mathrm{ov}=0.33$) and semiconvection ($\alpha_\mathrm{sc}=1$) as in \citet{2020ApJ...888L..12W} to avoid central helium ignition in the Hertzsprung gap \citep{2019A&A...625A.132S}, as the following Case~C behaves differently than Case~B mass transfer, and to better compare our results with \citet{ChenPhD}. We ran our models until central helium depletion and used them to model the evolution of the donor star. For the accretor star, we assume no evolution, which is justified for mass ratios away from unity, and interpolate between the ZAMS models to build a binary grid for each donor mass with mass ratios from 0.1 to 0.95 in steps of 0.05 and orbital periods from $10^{-0.5}\days \approx 0.3\days$ to $10^{3.5}\days \approx 3000\days$ in steps of 0.25\,dex.

For each combination of initial donor mass $M_\mathrm{1i}$, initial accretor mass $M_\mathrm{2i}$ and initial orbital period, we determine the Roche radius of the donor using the fit formula from \citet{1983ApJ...268..368E}. If the Roche radius is equal to the stellar radius of a hydrogen core-burning donor model (RLO Case~A), we calculate the post-RLO donor mass $M_\mathrm{1f}$ according to Schürmann et~al. (in prep., eq.~4). If the Roche radius is equal to the stellar radius of a hydrogen shell-burning model (RLO Case~B) or a helium-burning model (RLO Case~C), we use the helium core-mass of the donor as the post-RLO donor mass $M_\mathrm{1f}$.

Next we determine the logarithmic timescale-ratio $t$ (Eq.~\ref{eq:t}). The thermal timescale of the accretor star is given by Eq.~\ref{eq:kh} using the ZAMS values. The mass-transfer timescale can be calculated using Eq.~\ref{eq:md}. $\dot M_2$ is given by by $-\eps \dot M_1$, which in turn can be estimated by $\dot M_1 \approx (M_\mathrm{1f}-M_\mathrm{1i})/\tau_\mathrm{KH1}$. The mass transfer efficiency $\eps$ is assumed to be constant during the RLO and is a free parameter. Thus we find
\begin{equation}\label{eq:tapprox}
    t \approx \log \left( \frac{M_\mathrm{2i}^2}{M_\mathrm{1i}^2} \cdot \frac{R_\mathrm{1i} L_\mathrm{1i}}{R_\mathrm{2i} L_\mathrm{2i}} \cdot \eps \cdot \frac{M_\mathrm{1i}-M_\mathrm{1f}}{M_\mathrm{2i}} \right).
\end{equation}
For $t<0$, the accretor remains in thermal equilibrium, as discussed in Sect.~\ref{sec1}, and we describe its radius evolution during the RLO by linearly interpolating $\log R_\mathrm{2i}$ to $\log R_\mathrm{2f}$ between $M_\mathrm{2i}$ and the final secondary mass $M_\mathrm{2f}$. For $0<t<t_\mathrm{max}$ (Eq.~\ref{eq:grenze}), we determine the parameters $r$ and $m$ (Eq.~\ref{eq:Rfit} and~\ref{eq:Mfit}). We model the time-dependent accretor radius linearly from $\log R_\mathrm{2i}$ at $M_\mathrm{2i}$ to $\log R_\mathrm{max} = \log R_\mathrm{2i} + r$ at $M_{R=R_\mathrm{max}} = mM_\mathrm{2i}$, and from $\log R_\mathrm{max} = \log R_\mathrm{2i} + r$ to $\log R_\mathrm{2f}$ between $M_{R=R_\mathrm{max}} = mM_\mathrm{2i}$ and $M_\mathrm{2f}$.  We chose this piece-wise linearity in $\log R$ due to the rough piece-wise linear behaviour shown in Fig.~\ref{fig:test}. For $t>t_\mathrm{max}$, we assume that the accretor swells until it reaches the Hayashi lines. We model these by assuming a fixed effective temperature of $\log T_\mathrm{eff}/\K = 3.6$ and an additional luminosity induced by the gravitational restructuring of the star, given by
\begin{equation}
    L_\mathrm{gr} = \frac{1}{2} \frac{G M_\mathrm{2i} \dot M_\mathrm{2i}}{R_\mathrm{2i}}.
\end{equation}
In general, we use this to limit the accretor radius. We have thus derived a model for the accretor radius under accretion as a function of the current accretor mass.

Similar to the mass transfer efficiency $\eps$, the angular momentum budget of an RLO is not well understood. To describe it, we use the formalism of \citet{1997A&A...327..620S}, which allows an analytical analysis of certain angular momentum budgets. In addition to their parameters $\alpha$, the fraction of mass lost from the donor leaving the system with the specific orbital angular momentum of the donor, and $\beta$, the mass fraction lost from the donor leaving the system with the accretor's specific orbital angular momentum, we introduce $\eta$ for the material ejected with the specific orbital angular momentum of the binary system. These quantities are related as $\eps=1-\alpha-\beta-\eta$. \citet{1997A&A...327..620S} introduces a parameter $A$ to describe the enhancement of the angular momentum loss at the donor by spin-orbit coupling. We will use it as a general parameter to scale the angular momentum loss and also introduce $B$ and $H$ as scaling factors for the angular momentum loss similar to $A$. This leads the three angular momentum evolution exponents in \citet{1997A&A...327..620S} eqs.~(25) to~(27) taking the form
\begin{align}
    \mathcal A_w &= A \alpha, \\
    \mathcal B_w &= \frac{A\alpha + B\beta - H\eta}{1-\eps}, \\
    \mathcal C_w &= \frac{A\alpha\eps + B\beta/\eps - H\eta}{1-\eps}.
\end{align}

We use eq.~(28) from \citet{1997A&A...327..620S} to determine the system's semi-major axis as a function of the accretor mass, expressed by the mass ratio. With this, we use the formula of \citet{1983ApJ...268..368E} to calculate the Roche radius of the accretor $R_\mathrm{RL2}$ over the course of the RLO and eq.~(3.1) and~(3.2) from \citet{PabloPhD} to find the L2 equivalent-radius $R_\mathrm{L2}$\footnote{Technically, L2 and L3 flip when the mass ratio inverts. Therefore one has to evaluate eq.~(3.1) to find L3.}. Finally, we calculate the maximum of $\log(R_2/R_\mathrm{RL2})$ and $\log(R_2/R_\mathrm{L2})$ during the RLO as a function of initial mass ratio and initial orbital period for fixed donor mass, mass transfer efficiency and angular momentum budget. The sign of these tells us whether the accretor remains within its Roche lobe or whether contact or even a L2 overflow occurs.

\subsection{Results}\label{sec2:res}

\begin{figure}
    \centering
    \includegraphics[width=\hsize]{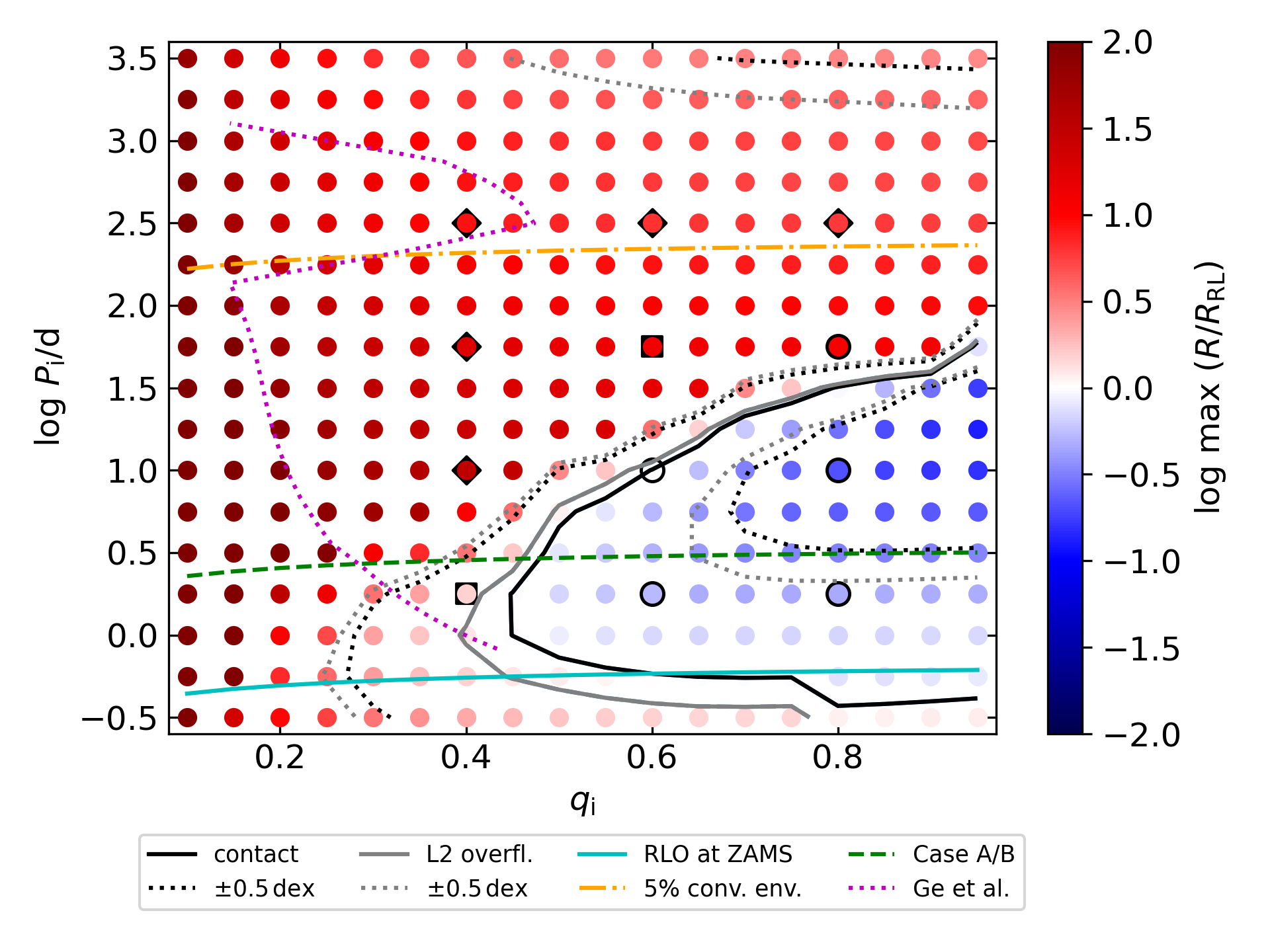}
    \includegraphics[width=\hsize]{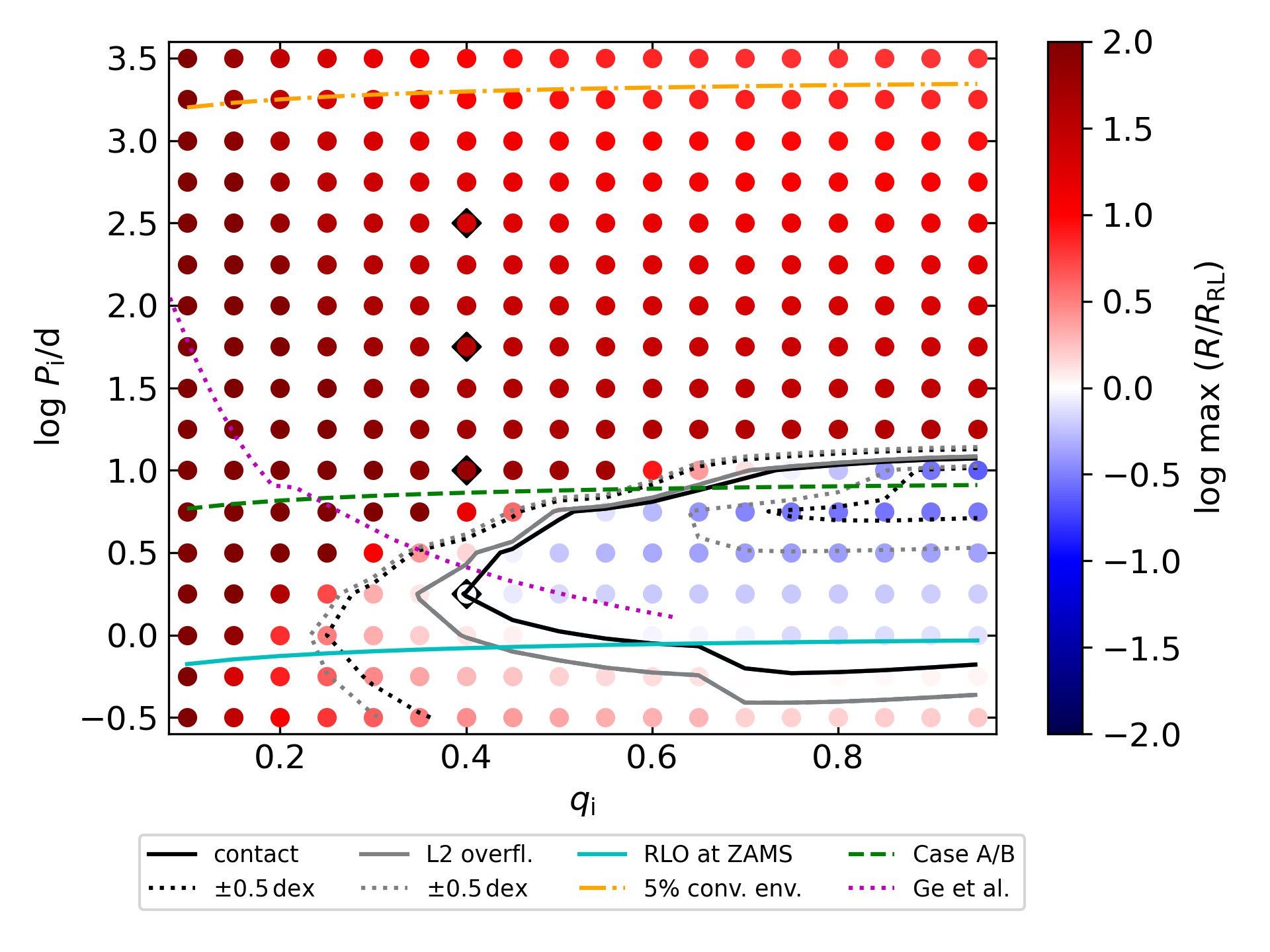}
    \caption{Maximum ratio of the accretor radius to its Roche radius over the course of the RLO as a function of initial mass and initial orbital period in our simulated binary systems. Red means that the accretor star swells to become larger than its Roche lobe, and blue indicates that it remains within its Roche lobe. The black line marks the boundary between these, and in grey we show the boundary for the L2 radius instead of the Roche radius. The dotted black and grey lines indicate a shift of $\pm0.5\,\mathrm{dex}$ in the two boundaries. The initial donor masses are $10\msol$ (top) and $30\msol$ (bottom), the mass transfer efficiency is 50\%, and we assume that the ejected material carries the specific orbital angular momentum of the accretor ($\alpha=\eta=0$, $\beta=-\eps$, $B=1$). We show the region where RLO at ZAMS is expected (cyan solid line), the boundary between Case~A and Case~B (green dashed line), the onset of the donor convective envelope (orange dot-dashed line), and the critical mass ratio according to \citet[pink dotted line]{2010ApJ...717..724G,2015ApJ...812...40G,2020ApJ...899..132G}. Systems marked with black symbols are those for which we have computed detailed models, see Table~\ref{tab:bin}. Black circles indicate stable swelling without contact formation, squares stand for stable swelling but L2 overflow, and diamonds indicate unstable swelling with L2 overflow.}
    \label{fig:10-30}
\end{figure}

In Fig.~\ref{fig:10-30} we show the maximum ratio of the accretor radius to its Roche radius over the course of the RLO for a $10\msol$ (top) and a $30\msol$ (bottom) donor for a mass transfer efficiency of 50\%, assuming that the ejected material carries the specific orbital angular momentum of the accretor, as a function of initial mass ratio and initial orbital period. Red dots indicate that the accretor is growing larger than its Roche lobe, and blue dots are for accretors that remain within their Roche lobe. In the $10\msol$-model, we find that for low initial orbital periods and mass ratios greater than about 0.5, the accretor avoids filling its Roche lobe and the system does not evolve to contact. For higher periods and lower mass ratios we get the opposite. Contact is also expected for very low orbital periods below about 0.3\,d, but this region is almost completely excluded due to the donor RLO at ZAMS. A similar pattern is observed for the L2 overflow, which is not unexpected since the L2-radius is not larger than about 30\% of the Roche radius.

We can explain why the accretors in low mass ratio systems tend to evolve to contact by using the thermal timescale of the accretor. The smaller the mass ratio, the smaller the accretor mass and luminosity, hence a larger accretor thermal timescale and a larger logarithmic timescale ratio $t$, since no changes have been made to the donor. The tendency for systems with large orbital periods to develop contact can be understood by the thermal timescale of the donor. A larger orbit implies a larger donor radius at the start of RLO, which implies a smaller donor thermal timescale and thus a larger mass transfer rate, resulting in a larger logarithmic timescale ratio $t$.

For the adopted angular momentum budget (material leaving the system carries the same specific angular momentum as the accretor), we find Case~A and Case~B systems that can avoid contact. The cyan line, which marks RLO at ZAMS, indicates the lower period limit for meaningful binary evolution. We expect the accretor not to overfill its Roche lobe in systems with orbital periods above 3000\,d, indicated by the dotted lines in the upper right-hand corner. However, this is hardly relevant as the donor is barely expected to grow large enough in radius to initiate a RLO as such high initial orbital period.

Moving to the $30\msol$ donor, we find that that the contact avoidance region is smaller compared to the $10\msol$ model. In particular, almost no Case~B system can avoid contact. It is a general trend that we observe that as the donor mass increases, a smaller number of systems avoid contact. We explain this with Fig.~\ref{fig:Rmax}. Higher donor masses imply higher accretor masses. For higher accretor masses, $t_\mathrm{max}$ approaches zero, shrinking the contact avoidance region.

\begin{figure}[h!]
    \includegraphics[width=\hsize]{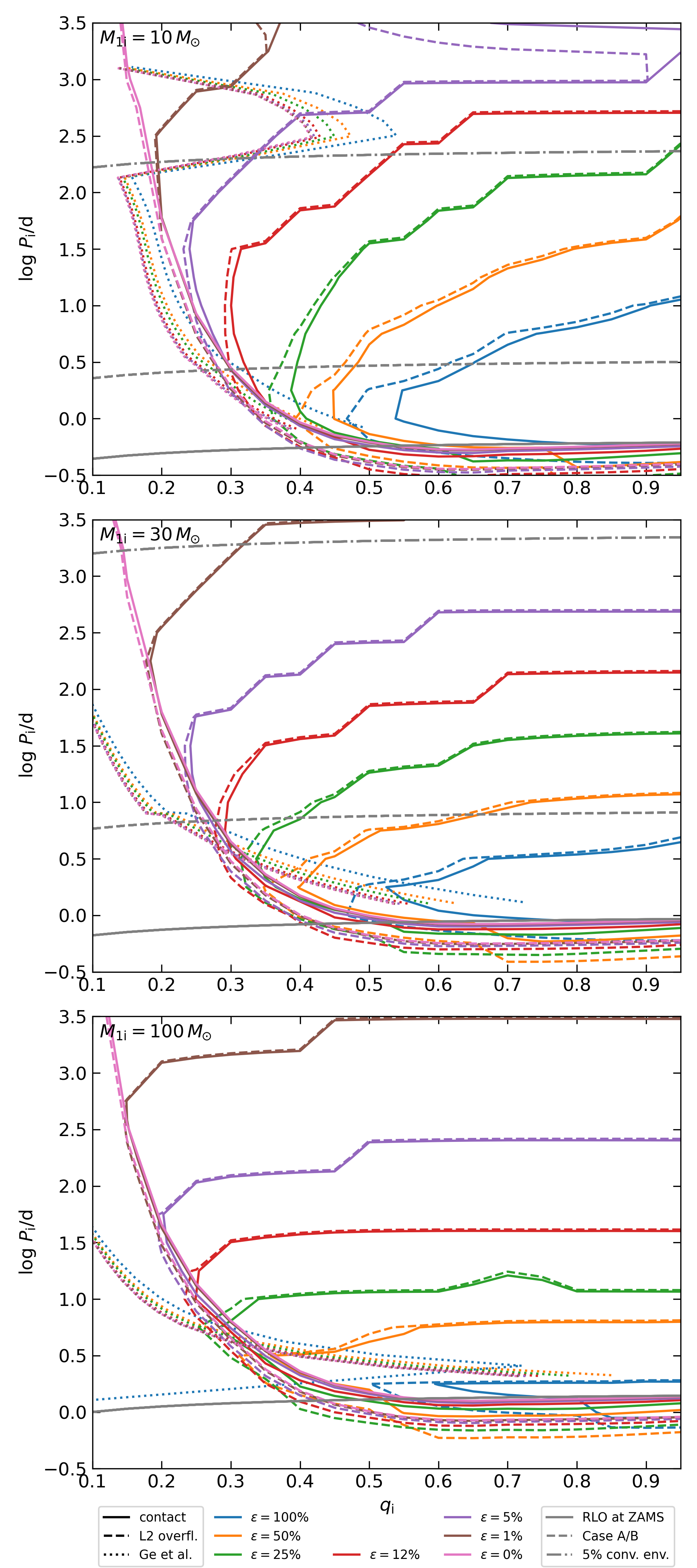}
    \caption{Boundary between the contact-developing and the contact-avoiding models for different initial donor masses (from top to bottom 10, 30, and $100\msol$) assuming the ejected material carries the accretor's specific orbital angular momentum (i.e. $\alpha=\eta=0$, $\beta=-\eps$, $B=1$). Colours indicate the assumed mass transfer efficiency. Dashed lines show the boundary of L2 overflow and dotted lines the critical mass ratio for dynamical timescale mass transfer derived from \citet{2010ApJ...717..724G,2015ApJ...812...40G,2020ApJ...899..132G}.}
    \label{fig:b1}
\end{figure}

In Fig.~\ref{fig:b1} we show only the boundary between contact and non-contact models, but for varying accretion efficiencies. At the highest efficiency, the contact avoidance region is the smallest and it grows with decreasing efficiency. This can be understood from Eq.~\ref{eq:tapprox}, where $t \propto \log\eps$. For the assumed angular momentum budget, we also find a limiting case given by completely non-conservative mass transfer (brown line). This line indicates the limit where contact on the left cannot be avoided due to the orbital evolution of the system.

Figs.~\ref{fig:b20} to~\ref{fig:h12} show the same as Fig.~\ref{fig:b1}, but with different assumptions about the amount of angular momentum of the ejected material. If it carries twice the specific orbital angular momentum of the accretor (Fig.~\ref{fig:b20} left), the main difference from the original case is that the boundary for unavoidable contact has moved to the right. If the ejected material carries the donor specific orbital angular momentum (Fig.~\ref{fig:a12} left), this boundary exists only towards small initial periods. The pattern remains that a higher mass transfer efficiency implies more contact systems. If twice this angular momentum is ejected (Fig.~\ref{fig:a12} right), for most mass transfer efficiencies at low donor mass, most Case~A mass transfers lead to contact. If the ejected material carries the binary's specific orbital angular momentum once or twice (Fig.~\ref{fig:h12}), a mixture of the above cases occurs. If no angular momentum is lost (Fig.~\ref{fig:b20} right), the patterns resemble the case where the ejected material carries the specific orbital angular momentum of the donor.

\subsection{Discussion}\label{sec2:dis}
In this Section we describe the uncertainties of our model (Sect.~\ref{sec2:uncert}), compare it with detailed binary models in Sect.~\ref{sec2:test}, apply our recipe to the WR stars in the SMC (Sect.~\ref{sec2:wr}), and compare our work with previous publications in Sect.~\ref{sec2:prev}.

\subsubsection{Uncertainties}\label{sec2:uncert}
We have already described the uncertainties in modelling the swelling of the accretor in Sect.~\ref{sec1:uncert}. Besides that, for the model of the binary system, the most important uncertainty is probably that we have neglected the nuclear evolution of the accretor. This assumption is only realistic for systems with mass ratios not close to unity, where the nuclear evolution of the more massive star is much faster than that of the companion. If this is not the case, stellar models predict that both the radius and luminosity of the star will have increased at the onset of accretion, and thus the thermal timescale of the accretor has decreased. This causes the logarithmic timescale ratio $t$ to be smaller given the same accretion rate. We therefore expect the accretor to expand less, which should increase the region of contact avoidance region in Fig.~\ref{fig:10-30} for mass ratios close to unity by shifting its upper boundary upwards.

The lack of nuclear evolution of the accretor also causes our models to avoid a reverse mass transfer and a mass transfer on post-main-sequence models. This is important for mass ratios very close to unity and/or Case~A systems. If both stars have similar masses, the accretor star can complete its central hydrogen burning during or after the RLO, which is expected to result in a merger \citep{2001A&A...369..939W,2022A&A...659A..98S}. For a Case~A RLO this is much more likely, even for mass ratios not very close to unity, since this type of mass transfer proceeds on the nuclear timescale \citep{1994A&A...290..119P,2001A&A...369..939W,2022A&A...659A..98S}. A comparison of our Fig.~\ref{fig:10-30} with the yellow shaded area of fig.~A.4 (left) of \citet{ChenPhD} suggests that at least half of the systems in the contact-avoiding region of Case~A should undergo this effect.

Our adopted values for semiconvection and overshooting avoid central helium ignition in the Hertzsprung-gap \citep{2019A&A...625A.132S,2020A&A...638A..55K,2022A&A...662A..56K}. These works and other recent studies such as \citet{2021A&A...653A.144H} favour models which avoid the red supergiant stage during core helium burning. Early central helium ignition would convert most of our Case~B systems to Case~C systems, if they reach the required radius, or lead to helium ignition during the RLO, which may cause its termination. The formation of blue loops starting from a red supergiant is unproblematic, as the radius evolution beyond the Hertzsprung gap is slow enough to make helium ignition during RLO is unlikely.

In this work we have restricted ourselves to angular momentum budgets that can be described by analytical formulae, because these were easy to implement. In a real binary the ejection of material from the binary may be a complex hydrodynamical process and thus the angular momentum budget could be more complex than the formalism of \citet{1997A&A...327..620S}. However, we have analysed the limiting case where no angular momentum leaves the system (Fig.~\ref{fig:b20} right). On the other hand, if the angular momentum loss is much larger than assumed here, say of the order of an L2 overflow, we expect the system to undergo a rapid merger.

Our results were found using stellar models with SMC metallicity. In Sect.~\ref{sec1:prev}, we found evidence that with higher metallicity, the accretor swells more. A larger maximum accretor radius means that the system is more likely to evolve into contact, and thus the contact-avoidance regions in Fig.~\ref{fig:b1} should become smaller. This means that at larger metallicities the products of stable mass transfer could be less, and common envelope or merger products could be more likely.


\subsubsection{Comparison with detailed binary models}\label{sec2:test}

In Fig.~\ref{fig:test} we have indicated by $+$-symbols the maximum of the mass-radius curve when using Eq.~\ref{eq:tapprox} with the parameters of the detailed models at the onset of RLO. They agree well with the $\times$-symbols, which are based on the actual mass transfer rate of the detailed models, except for models \#1 (blue), \#7 (brown) and \#11 (yellow). For the latter two this is unproblematic because the binary is very wide and the maximum radius is either much larger (\#7) or much smaller (\#11) than the Roche radius.

In Fig.~\ref{fig:10-30} we have shown the initial-mass--initial-period combination for which we have computed detailed MESA binary models with black symbols. Whether they evolve into contact or avoid it can be predicted well by our method. Systems that remain in thermal equilibrium or swell stably but do not overfill their Roche lobe are marked with circles. Indeed, models \#5, \#6, \#9, and  \#10 are in the contact avoidance region. However, model \#6 is near the boundary and model \#11 deviates from our prediction, likely because the accretor in our simplified models does not undergo nuclear evolution before RLO, see Sect.~\ref{sec2:uncert}. Models \#1 and \#7 are computed to swell stably but still overfill L2 (squares), which is indeed confirmed in Fig.~\ref{fig:10-30}. Unstable swelling and subsequent L2 overflow is observed in models \#2 and \#3 and expected to occur in models \#4, \#8 and \#12, all marked by diamonds and all in the contact forming region. For the $30\msol$-donor, models \#14, \#15, and \#16 behave as predicted. Model \#13 swells unstably and undergoes L2 overflow, but our recipe predicts that it will just fill its Roche lobe. As the function we found for $r$ (Eq.~\ref{eq:Rfit}) is quite steep, especially at high masses (see Fig.~\ref{fig:Rfit}), it is not unexpected that such mismatches occur at the boundaries.

Finally, Fig.~\ref{fig:10-30} shows that our assumption that unstably swelling models can be assumed to reach the Hayashi lines is reasonable. Models \#2 and \#3 swell to a radius similar to the position of the $+$-symbols. Also, only a small mass accretion ($0.5\msol$ and $1\msol$ respectively) is required.

\subsubsection{Comparison with the SMC binary WR stars}\label{sec2:wr}

It is instructive to compare our results with the WR stars of the SMC. Indeed, it is proposed that WR stars can form by binary interaction \citep[e.g.][]{2020A&A...634A..79S,2022A&A...667A..58P}. The SMC contains twelve WRs, five of which are known binaries \citep[][Schootemeijer et al. subm.]{2003MNRAS.338..360F,2004A&A...416..291F,2016A&A...591A..22S}. One of them, AB\,5, is a double WR star, which makes it unsuitable for the following analyses. Following \citet{2018A&A...611A..75S}, we can add the initial mass ratio and initial orbital period of the remaining WR+O systems (AB\,3, AB\,6, AB\,7, and AB\,8) to diagrams like Fig.~\ref{fig:b1}. We can calculate the initial orbital period from the observed orbital period, if the current and initial mass ratios are known and the angular momentum budget is fixed \citep[Sect.~\ref{sec2:met} and][]{1997A&A...327..620S}. To determine the current mass ratio, we can rely on radial velocity variations \citep[][Schootemeijer et al. subm.]{2003MNRAS.338..360F,2016A&A...591A..22S,2018A&A...616A.103S} or on mass estimates of the two stars. \citet{2016A&A...591A..22S,2018A&A...616A.103S} estimate the mass of the O~star in two ways. One mass estimate is derived from the spectral type, and the other is derived from the surface gravity. Unfortunately, these estimates typically have significant uncertainties. The WR masses from \citet{2016A&A...591A..22S,2018A&A...616A.103S}, based on the mass-luminosity relation of \citet{2011A&A...535A..56G} agree well with those from \citet{2018A&A...611A..75S}. 
Thus for each system we can adopt two of the three observed properties (mass ratio, WR~mass, O-star mass) and calculate the third. Furthermore, we can use the WR masses together with their hydrogen surface abundance to estimate the initial masses of the WR progenitor 
using the models of \citet{2019A&A...625A.132S} with overshooting and semiconvection as suggested by \citet{2021A&A...653A.144H}. Finally, assuming a mass transfer efficiency $\eps$, we can calculate the initial O~star mass $M_\mathrm{2i}=M_\mathrm{O}-\eps\cdot(M_\mathrm{1i}-M_\mathrm{WR})$ and find the initial mass ratio and initial orbital period. We have summarised the adopted values in Table~\ref{tab:wr}.

\begin{table*}[]
    \centering
    \caption{Mass estimates and orbital parameters of the four WR+O~systems in the SMC. The superscript indicates the method on which the estimate is based (\textDelta RV = radial velocity variation, lum = Luminosity, $\mathrm{d}X/\mathrm{d}Q$ = hydrogen gradient, SpT = spectral type, gr = surface gravity)}
    \label{tab:wr}
    \begin{tabular}{ccccc} \hline\hline
        & AB\,3 & AB\,6 & AB\,7 & AB\,8 \\ \hline
        $P_\mathrm{orb}/\days$ & $10.053(5)$\tablefootmark{(f)} & $6.5384(4)$\tablefootmark{(g)} & $19.5600(5)$\tablefootmark{(h)} & $16.633(9)$\tablefootmark{(i)} \\
        $q^\text{\textDelta RV}$ & --\tablefootmark{(f)} & $2.23(9)$\tablefootmark{(g)} & $1.94(6)$\tablefootmark{(h)} & $2.85(20)$\tablefootmark{(i)} \\
        $M_\mathrm{WR}^\mathrm{lum}/\msol$\tablefootmark{(a)} & $29^{+2}_{-2}$ & $26^{+7}_{-5}$ & $37^{+6}_{-5}$ & $40^{+7}_{-6}$ \\
        $M_\mathrm{WR}^{\mathrm{d}X/\mathrm{d}Q}/\msol$\tablefootmark{(b)} & $28$ & $20\dots30$\tablefootmark{(c)} & $34$ & -- \\
        $X_\mathrm{H,WR}$\tablefootmark{(d)} & $0.25(5)$ & $0.25(5)$ & $0.15(5)$ & $0^{+0.15}$ \\
        $M_\mathrm{O}^\mathrm{SpT}/\msol$\tablefootmark{(d)} & $20^{+20}_{-5}$ & $41^{+10}_{-10}$ & $44^{+26}_{-9}$ & $61^{+14}_{-25}$ \\
        $M_\mathrm{O}^\mathrm{gr}/\msol$\tablefootmark{(d)} & $13^{+70}_{-10}$ & $61^{+60}_{-30}$ & $30^{+55}_{-19}$ & $70^{+210}_{-52}$ \\
        $M_\mathrm{1i}^\text{lum}/\msol$\tablefootmark{(e)} & $50^{+5}_{-5}$ & $47^{+10}_{-7}$ & $65^{+10}_{-10}$ & $80^{+15}_{-15}$ \\
        $M_\mathrm{1i}^\text{SpT $\wedge$ \textDelta RV}/\msol$\tablefootmark{(e)} & -- & $35^{+5}_{-5}$ & $45^{+20}_{-7}$ & $47^{+8}_{-15}$ \\
        $M_\mathrm{1i}^\text{gr $\wedge$ \textDelta RV}/\msol$\tablefootmark{(e)} & -- & $50^{+40}_{-20}$ & $32^{+43}_{-15}$ & $50^{+?}_{-30}$ \\
        \hline
    \end{tabular}
    \tablefoot{The superscripts indicate the method used to estimate the masses. \tablefoottext{a}{\citet{2016A&A...591A..22S,2018A&A...616A.103S} using the mass-luminosity relation from \citet{2011A&A...535A..56G}.} \tablefoottext{b}{\citet{2018A&A...611A..75S}, since their method requires $X_\mathrm{H,WR}>0$, no mass could be determined for AB\,8.} \tablefoottext{c}{Schootemeijer (priv. comm.)} \tablefoottext{d}{\citet{2016A&A...591A..22S,2018A&A...616A.103S}} \tablefoottext{e}{Based on $M_\mathrm{WR}^\mathrm{lum}/\msol$, $X_\mathrm{H,WR}$ and the profiles of \citet{2019A&A...625A.132S}. The mass for AB\,8 is a lower limit.} \tablefoottext{f}{\citet{2003MNRAS.338..360F}, no mass ratio could be determined.} \tablefoottext{g}{\citet{2018A&A...616A.103S}} \tablefoottext{h}{\citet{2002MNRAS.333..347N}} \tablefoottext{i}{\citet{2001MNRAS.324...33B,2005ApJ...628..953S}}}
\end{table*}

\begin{figure}
    \centering
    \includegraphics[width=\hsize]{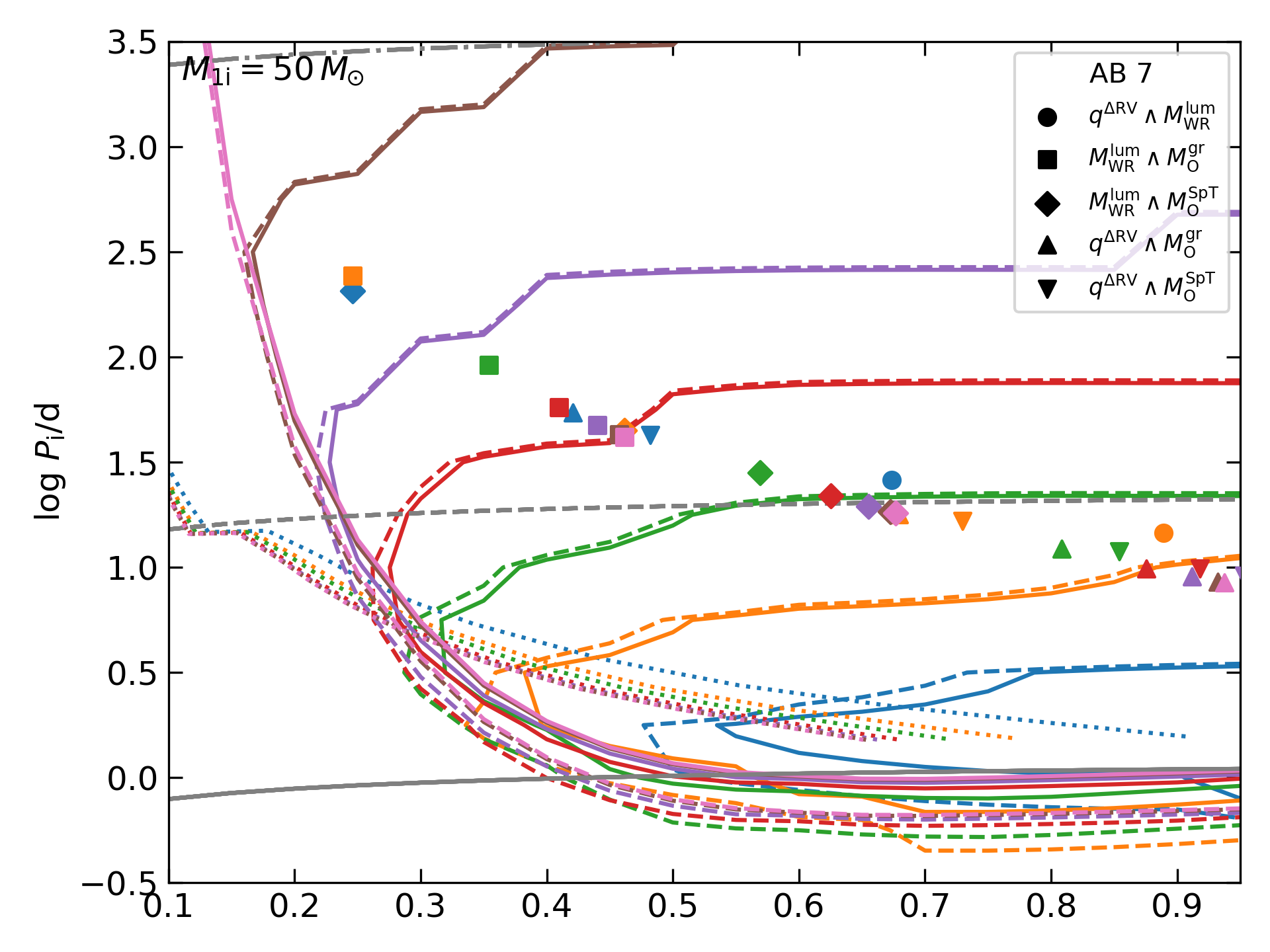}
    \includegraphics[width=\hsize]{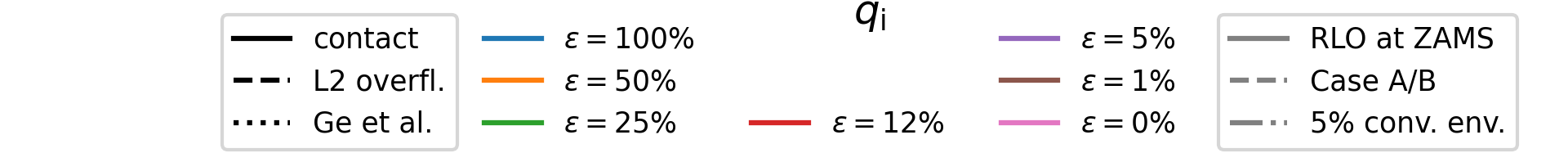}
    \caption{Same as Fig.~\ref{fig:b1}, but for a $50\msol$ donor star. The symbols indicate possible initial configurations for AB\,7. The shapes of the symbols indicate whether we have estimated the O~star mass with the mass ratio from radial velocity variations and the WR mass from its luminosity (circle), estimated the mass ratio with the surface-gravity mass of the O~star and the luminosity mass of the WR star (squares), with the spectral-type mass of the O~star and the luminosity mass of the WR star (diamonds), estimated the WR mass the the mass ratio and the surface-gravity mass of the O~star (triangles up), or the the mass ratio and the spectral-type mass (triangles down). See also Table~\ref{tab:wr}.}
    \label{fig:wr70}
\end{figure}

We show the resulting initial configurations of AB\,7 in Fig.~\ref{fig:wr70}. We find the initial donor mass of this system to be about $50\msol$ for all three estimates of the WR~mass. Based on the five different mass estimates for the WR and the O~star, we have placed the system several times in the diagram. We also vary the assumed mass transfer efficiency (colour). Comparing the proposed initial parameters of the systems with the corresponding contact boundary, we find that some values of the mass transfer efficiency lead to unrealistic results.

The case of conservative evolution (blue) places the system into the contact forming side of the diagram for all five methods. This means that under this assumption the system must have experienced contact or L2 overflow, which we have argued leads to a merger and not to the stripping of the envelope of the WR progenitor. On the other hand, only initial mass rations below unity are meaningful. This is fulfilled for all five methods except the one based on the WR~mass derived from the mass-luminosity relation and the spectroscopic mass ratio, for which we find initial mass ratios above unity for mass transfer efficiencies $>25\%$. All other methods yield initial configurations for the fully non-conservative case (pink) that are on the contract-avoiding side.

A closer inspection of the diagram reveals that for $\eps>50\%$ (orange) there are only unrealistic solutions and for $\eps<5\%$ (purple) there are only realistic ones. For $\eps=12\%$ (red), the mass estimate based on the mass-luminosity relation and the surface gravity is in the merger region and for $\eps=25\%$ (green) also the estimate with mass-luminosity relation and spectral type. 
This suggests that the mass transfer efficiency for AB\,7 was less than $50\%$, may be as low as $5\%$. A similar analysis for AB\,3 yields $\eps<1\dots5\%$, for AB\,6 $\eps<50\%$, and for AB\,8 $\eps<25\%$ (Fig.~\ref{fig:wr_other}).

Assuming other angular momentum budgets give similar results. In general, the ejection of more specific orbital angular momentum requires a lower mass transfer efficiency to obtain realistic initial configurations. Furthermore, we find that in almost all scenarios it was a Case~A RLO that formed AB\,6, AB\,7, and AB\,8 (initial configuration below the grey dashed line) while AB\,3 was likely formed in Case~B. This fits to the fact that we found it to have the lowest mass transfer efficiency of the analysed systems, as in close binary systems tides are expected to increase the mass transfer efficiency \citep[e.g.][]{2022A&A...659A..98S}. Also all four systems are stable according to the criterion of \citet{2010ApJ...717..724G,2015ApJ...812...40G,2020ApJ...899..132G} (initial configurations above the dotted lines). Unfortunately, the discrepancy between the different methods to estimate the stellar masses and the large errors of some of them makes a final judgement difficult. More constraining observations would be desired.

\subsubsection{Comparison with previous work}\label{sec2:prev}

The question under what conditions an RLO will lead to a stripped star and avoid a merger or a common envelope has been addressed by many authors. The classical approach is to compare the mass-radius indices of the donor and the Roche lobe. This is equivalent to asking whether the donor radius shrinks faster or slower under mass loss than its Roche radius. The most recent work on this topic is \citet{2010ApJ...717..724G,2015ApJ...812...40G,2020ApJ...899..132G}, who calculate mass-radius exponents $\tilde\zeta_\mathrm{ad}$ for donor stars of all evolutionary phases and also give critical mass ratios for conservative evolution. From their mass-radius exponents, we have derived critical mass ratios for all mass transfer efficiencies using eq.~(62) of \citet{1997A&A...327..620S}\footnote{Eq.~(30) (and eq.~(31)), which needs to be evaluated here, has a typo. The last term should have $\eps q$ in the numerator instead of $q$.}, as suggested by the authors, and plotted them as pink lines in Fig.~\ref{fig:10-30} and as dotted lines in Figs.~\ref{fig:b1} and~\ref{fig:b20} to~\ref{fig:h12}. They indicate that dynamical mass transfer is initiated to their left, probably leading to a merger or common envelope. In most cases, the critical mass ratios lie within the regions where we predict contact and L2 overflow to occur. Dynamical mass transfer only affects contact-avoiding systems with low accretion efficiencies or with high donor mass systems and short orbital periods. This means that our criterion -- the swelling of the accretor and subsequent contact and L2 overflow -- is stronger in deciding for or against a stable RLO.

\citet{2001A&A...369..939W} approach the occurrence of contact systems by computing a small grid of detailed binary models assuming conservative mass transfer. They distinguish three contact formation mechanisms. Their $q$-contact (contact due to higher mass transfer rates and/or larger thermal timescales of the secondary) is similar to what we observe in our grid. We cannot model what \citet{2001A&A...369..939W} call delayed contact, because our mass transfer rate is not resolved in time. However, we observe in our models, that in close and very unequal systems the secondary reaches its maximum radius at lower mass ($M_{R=R_\mathrm{max}} = m M_\mathrm{2i}$) than in wide and more equal systems. Finally, premature and reverse contact does not occur in our study, as we do not assume nuclear evolution of the secondary. We can compare fig.~12 of \citet{2001A&A...369..939W} with our Fig.~\ref{fig:b1} (top, conservative case). While they are qualitatively similar, \citet{2001A&A...369..939W} have a larger Case~B region with contract avoidance and a smaller corresponding region with Case~A region. This could come from the fact that although fast Case~A and Case~B take place on a thermal timescale, this is only the order of magnitude. Their duration in detailed binary models differs by a factor of a few. In agreement with them, we find the shrinking of the contact avoidance regions and the increasing dominance of Case~A with mass.

\citet{2020A&A...638A..39L}, \citet{2020ApJ...888L..12W} and \citet{2022A&A...659A..98S} use an energy criterion to determine the stability of the RLO. If the combined luminosity of the two stars is large enough to unbind the unaccreted material from the system, a merger will be avoided. They determine the amount of accretion by allowing the accretor to accrete until it reaches critical rotation. This leads to accretion efficiencies of less than 5\% for a Case~B mass transfer and up to 50\% for in Case~A \citep{2022A&A...659A..98S}. Although these detailed models would also indicate a merger by an L2 overflow, the accretion efficiencies are low enough that this only happens at very low initial orbital periods. Thus its in general the energy criterion which decides for or against a stable RLO. We compare our Fig.~\ref{fig:b1} with Fig.~A.4 (right) and~A.9 (right) of \citet{ChenPhD} because they have the same donor mass, metallicity and angular momentum budget. While for the $10\msol$-models \citet{ChenPhD} finds the region of stable mass transfer to take on a triangular shape between 5 and 100\,d up to a mass ratio of about 0.65 and a few Case~A systems up to a mass ratio of 0.8 to avoid contact, we find a much larger area if the accretion efficiency is below 50\%. Our shape of the contact-avoiding regions is also very different, especially since we find one but \citet{ChenPhD} find two separate regions. Its upper limit is given by the onset of convection in the donor. To allow the same number of systems to survive, we would have to set the mass transfer efficiency to about 100\%, which would however be centred on systems with close orbits. Comparing the $30\msol$-models, we see that our contact-avoiding regions have shrunk, but in \citet{ChenPhD} they have increased by a large amount. Almost all Case~B systems and half of the Case~A systems avoid contact. To allow the same number of systems to survive, we would have to set our mass transfer efficiency to about 5\%.

\citet{2015ApJ...805...20S} use fixed critical mass ratios for each Case~A and~B to decide as the merger criterion, but limit the accretion by the thermal timescale of the accretor \citep{hurley02}. In the language of our work, $\eps=1$ if $t<1$, and if $t$ would be greater than 1, $\eps$ is adjusted so that $t=1$. For a $10\msol$-donor this leads to fig.~1 of \citet{2015ApJ...805...20S}, where Case~A yields a stripped donor star for mass ratios greater than 0.56, not so different from our work, but their Case~B is split into a conservative case for high mass ratios and low orbital periods, and a highly non-conservative case otherwise. The line between these cases roughly corresponds to our line between contact and contact avoidance, but differs because our $t_\mathrm{max}$ is mass-dependent and we let the accretor swell. For higher donor masses \citet{2015ApJ...805...20S} find that the region of conservative mass transfer shrinks slightly (as does our contact-avoidance region), but the dividing line becomes an extended transition region (their $20\msol$ models in fig.~19).

\section{Conclusion and Outlook}\label{sec:concl}

Our work sheds new light on the question in which part of the initial binary parameter space the merging of the two stars can be avoided during their first mass transfer phase. When assuming a fixed mass transfer efficiency, the answer to this question depends on the swelling of the mass accreting star, and on the evolution of the orbital separation during the mass transfer. The former can be well computed by detailed binary evolution models, for which, however, it is difficult to scan through the rather unconstrained accretion and angular momentum loss efficiencies. Therefore, we develop a theoretical framework in which we first derive analytic approximations for the radius evolution of accreting main-sequence stars based on detailed models (Eqs.~\ref{eq:Rfit} and~\ref{eq:Mfit}), which we then use to predict the part of the initial binary parameter space in which merging can be avoided, as function of the chosen accretion efficiency and angular momentum loss parameter. 

We derived regions in the initial-mass-ratio--initial-orbital-period--plane in which the binary models evolve into L2-overflow, potentially leading to a merger, and the regions where the binaries avoid contact and develop a fully stripped donor with a main-sequence companion. We find that only very few models evolve into contact and avoid L2-overflow. We have tested and compared our binary evolution approach with detailed binary evolution models, for which we find reasonable agreement, but we find rather significant differences compared to simple  merger criteria often used in rapid binary evolution calculations. Our models predict a larger fraction of binaries to merge compared with most previous studies, even at the low metallicity of the Small Magellanic Cloud. For larger metallicities, we expect the mass gainers to swell more, which would result in more mergers.


We have applied our results to interpret the observations of the WR+O stars observed in the Small Magellanic Cloud. We found that the mass transfer process which has produced these binaries must have been inefficient, with a mass transfer efficiency of 50\% or less, and as low as 1\%  for one particular system.

We have developed a fast and powerful method to determine the outcome of a Roche-lobe overflow, which provides stronger constraints than the classical approach of comparing mass-radius exponents. We found that while the angular momentum loss parameter is not unimportant, the mass transfer efficiency is the more influential parameter. Our approach is suitable to be used in rapid population synthesis calculations, for which it is possible to avoid several of the simplifications made here, in particular neglecting the nuclear evolution of the mass gainer. In a forthcoming paper we will use our recipe in a rapid binary population synthesis of the post-mass transfer massive star population of the SMC.

\begin{acknowledgements}
 The authors would like to thank Pablo Marchant for allowing us to use his MESA framework, and the members of the Bonn Stellar Physics Group for the fruitful discussions.
\end{acknowledgements}

\bibliographystyle{aa}
\bibliography{bib}


\appendix
\onecolumn
\section{Hertzsprung-Russell diagrams of accreting stellar models}

\begin{figure*}[h]
    \includegraphics[width=0.5\hsize]{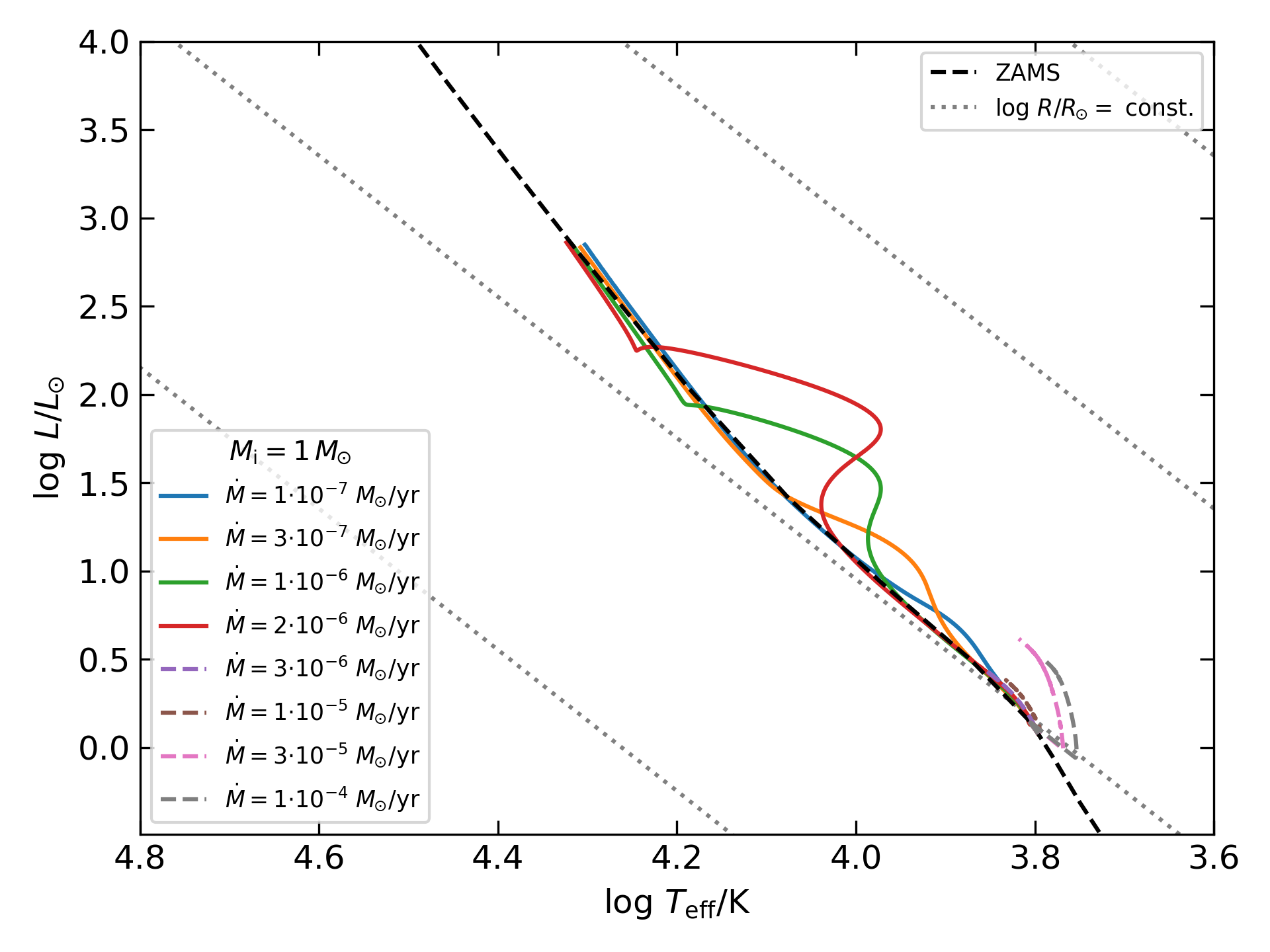}
    \includegraphics[width=0.5\hsize]{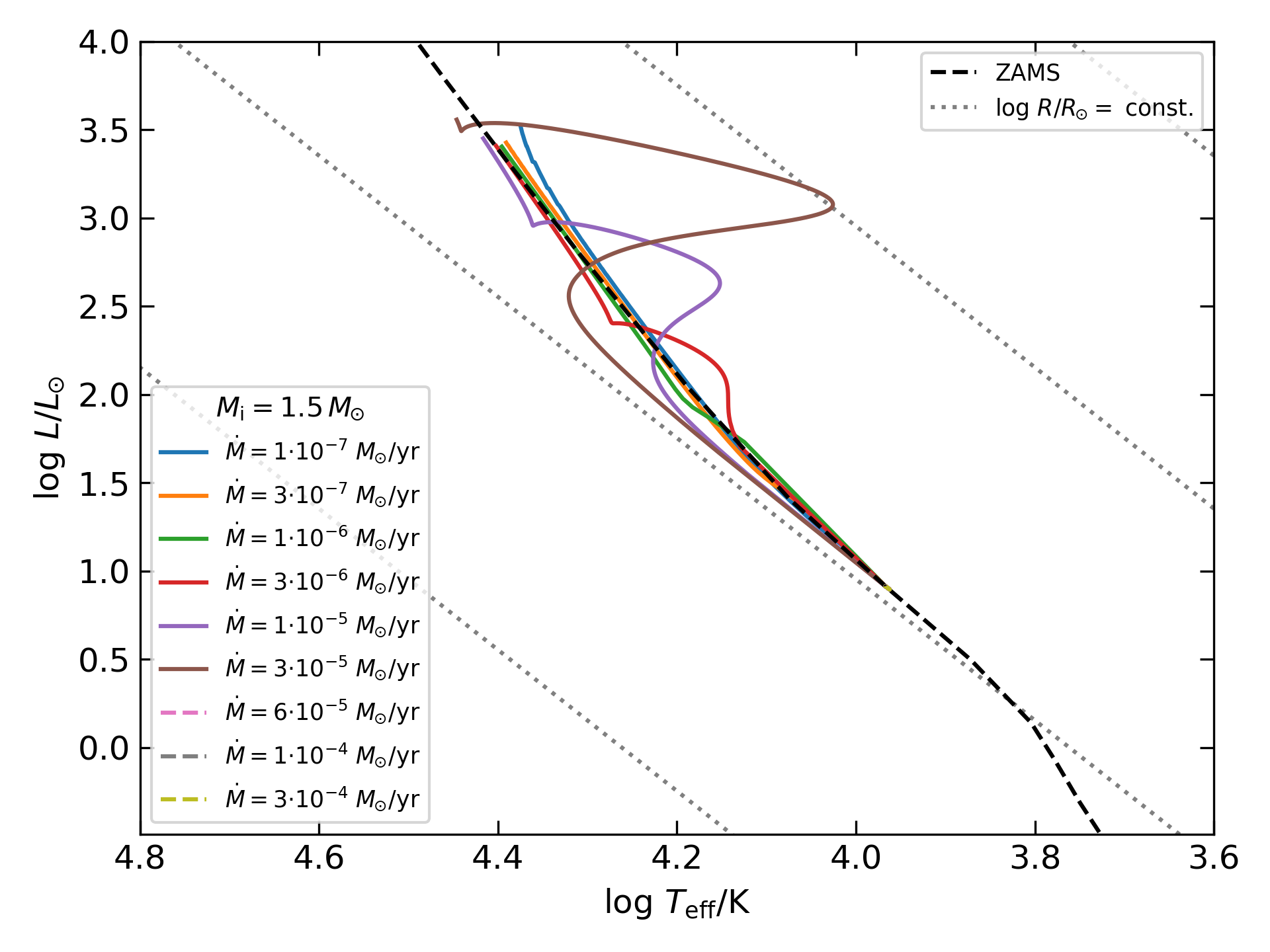}
    \includegraphics[width=0.5\hsize]{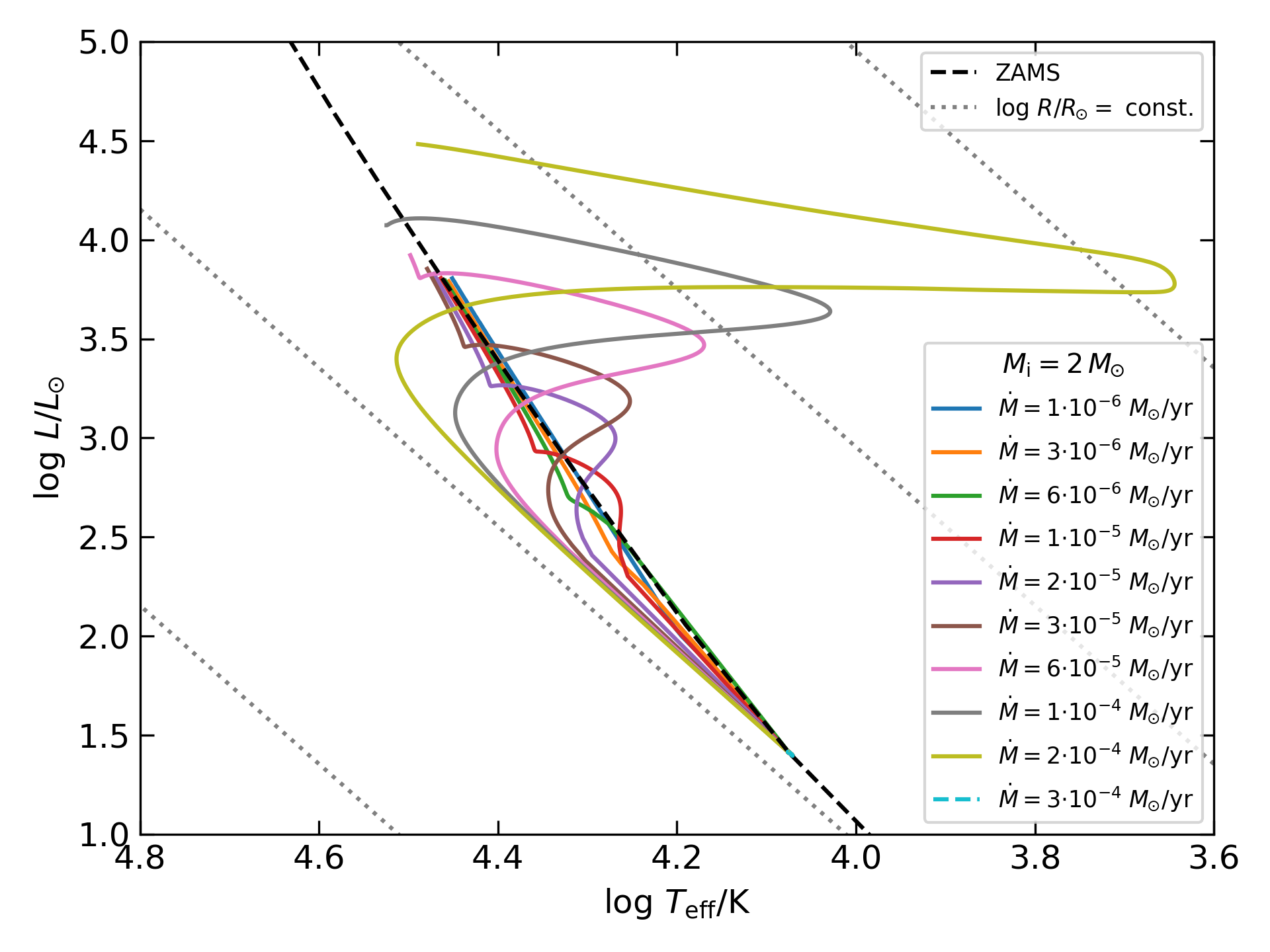}
    \includegraphics[width=0.5\hsize]{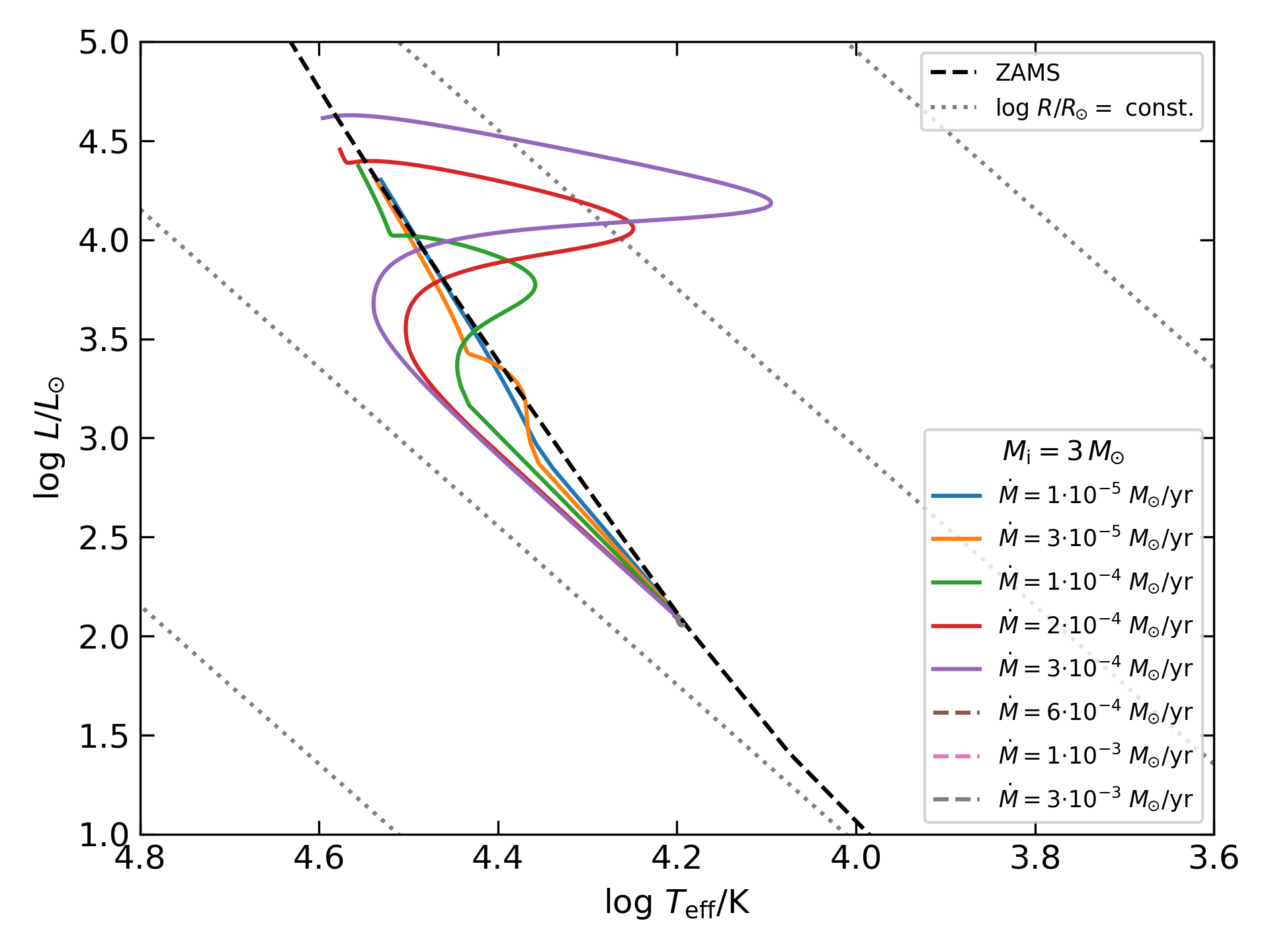}
    \includegraphics[width=0.5\hsize]{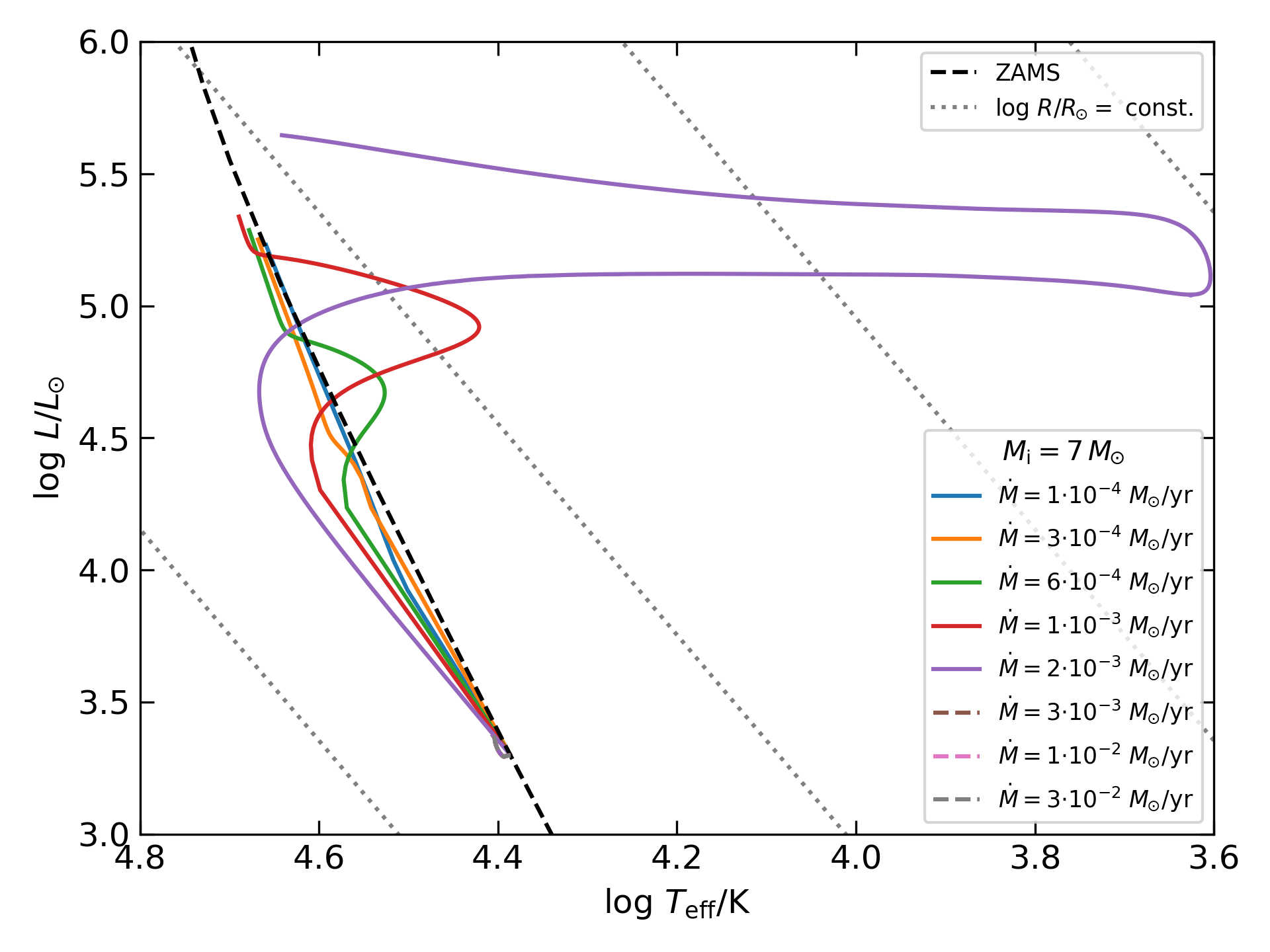}
    \includegraphics[width=0.5\hsize]{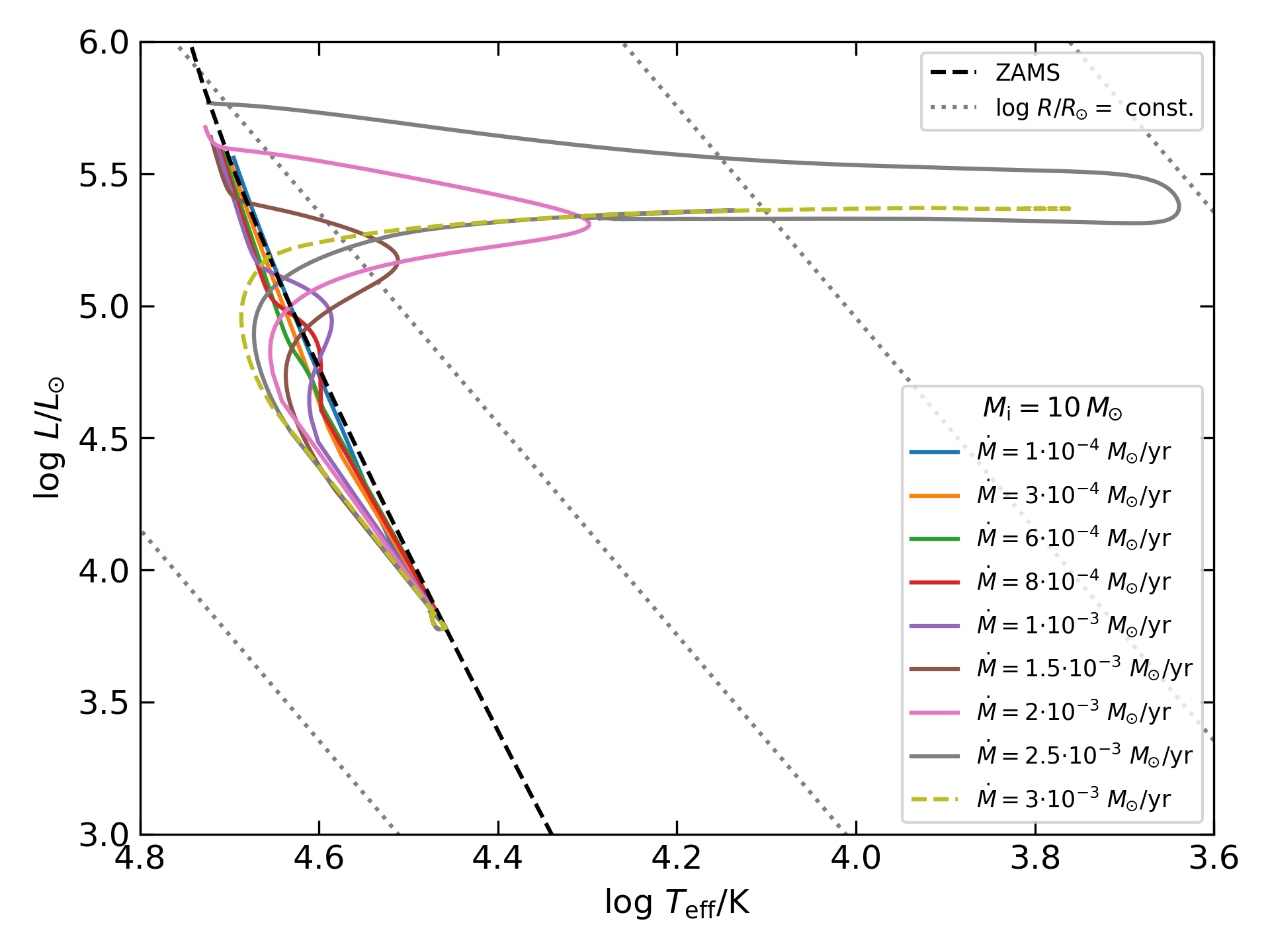}
    \caption{HRDs of our $1\msol$-, $1.5\msol$-, $2\msol$-, $3\msol$-, $7\msol$-, and $10\msol$-models for different accretion rates (indicated by colour) together with the final mass of the models.}
    \label{fig:hrd1}
\end{figure*}

\begin{figure*}[h]
    \includegraphics[width=0.5\hsize]{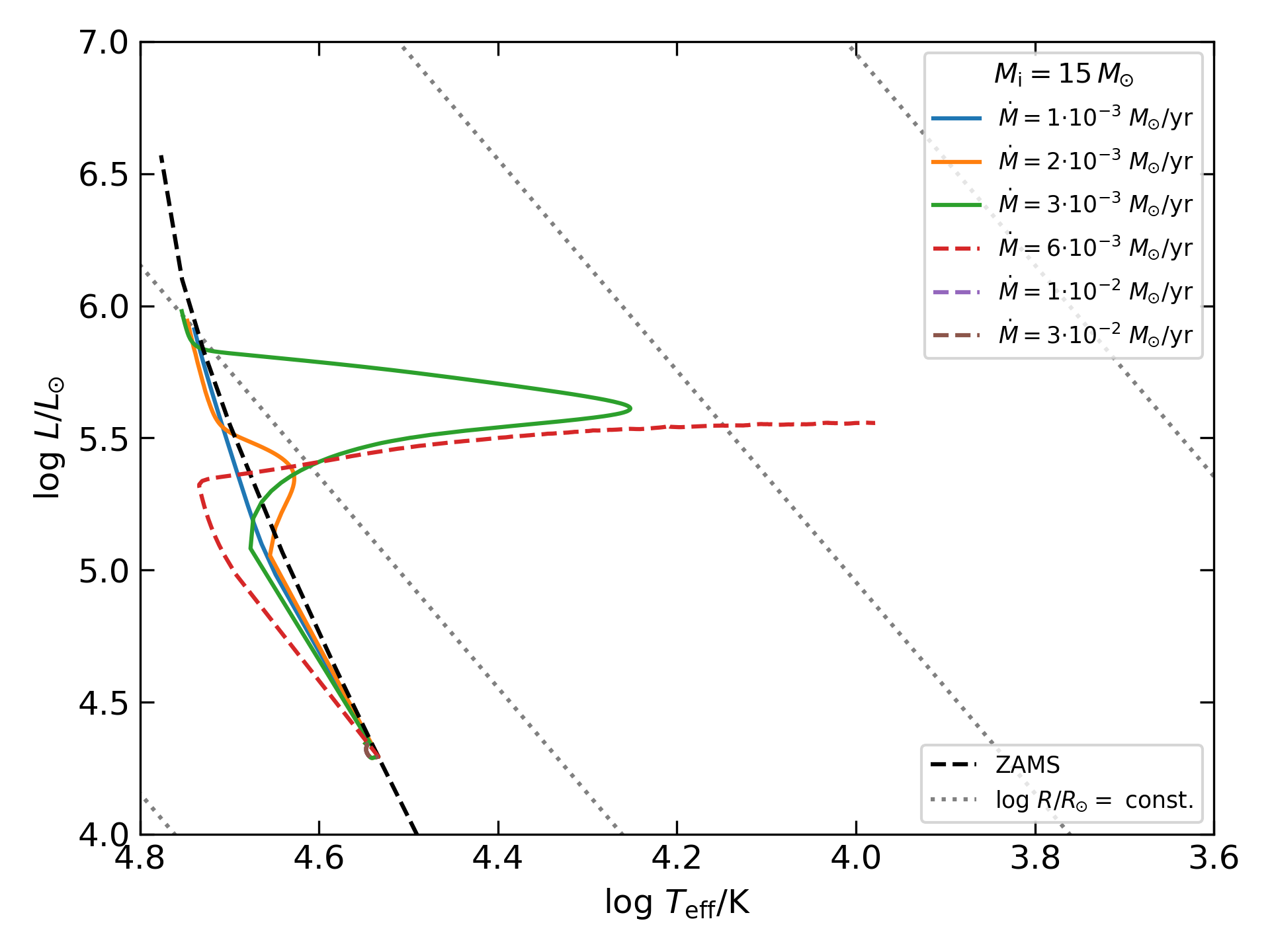}
    \includegraphics[width=0.5\hsize]{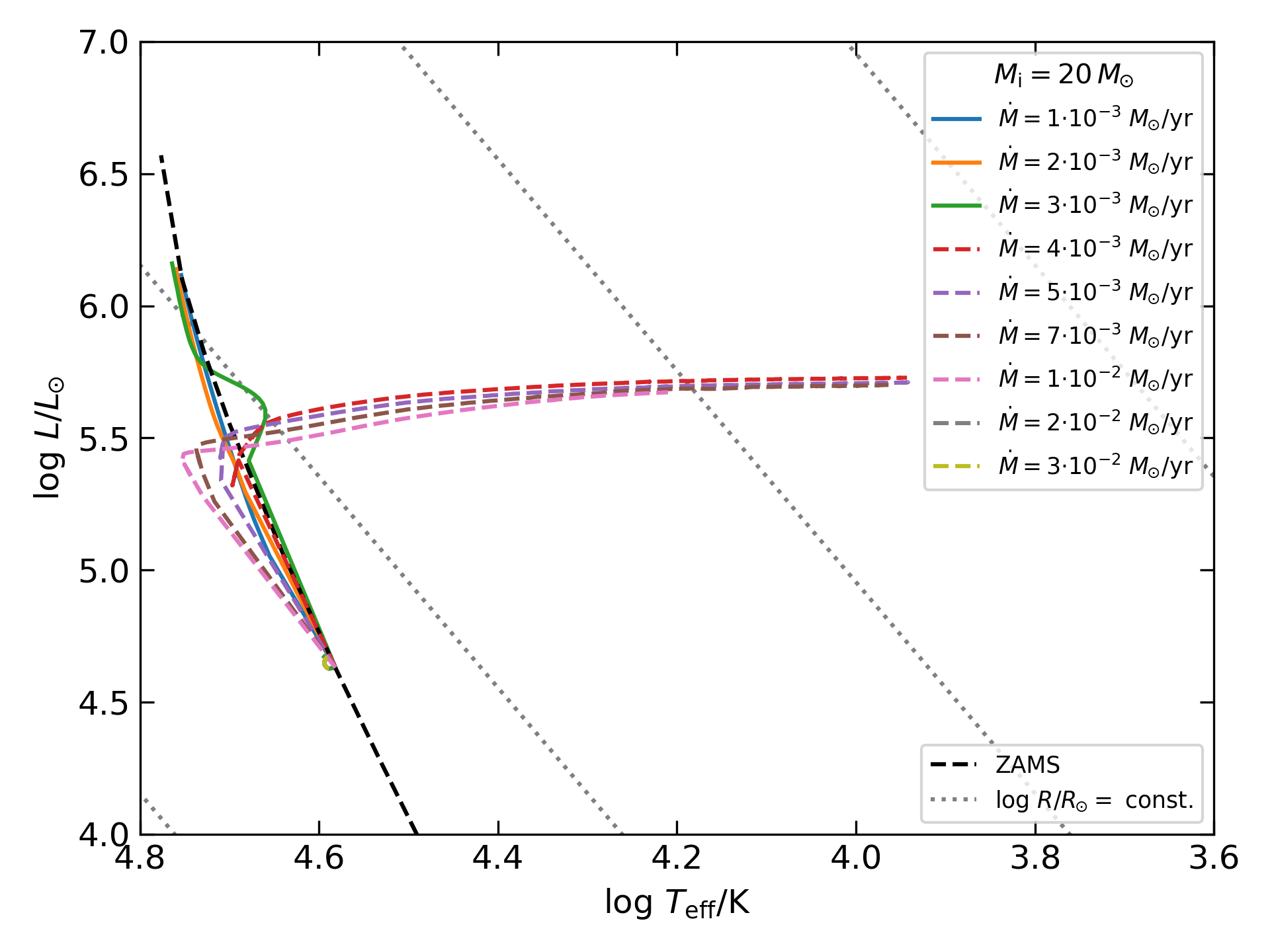}
    \includegraphics[width=0.5\hsize]{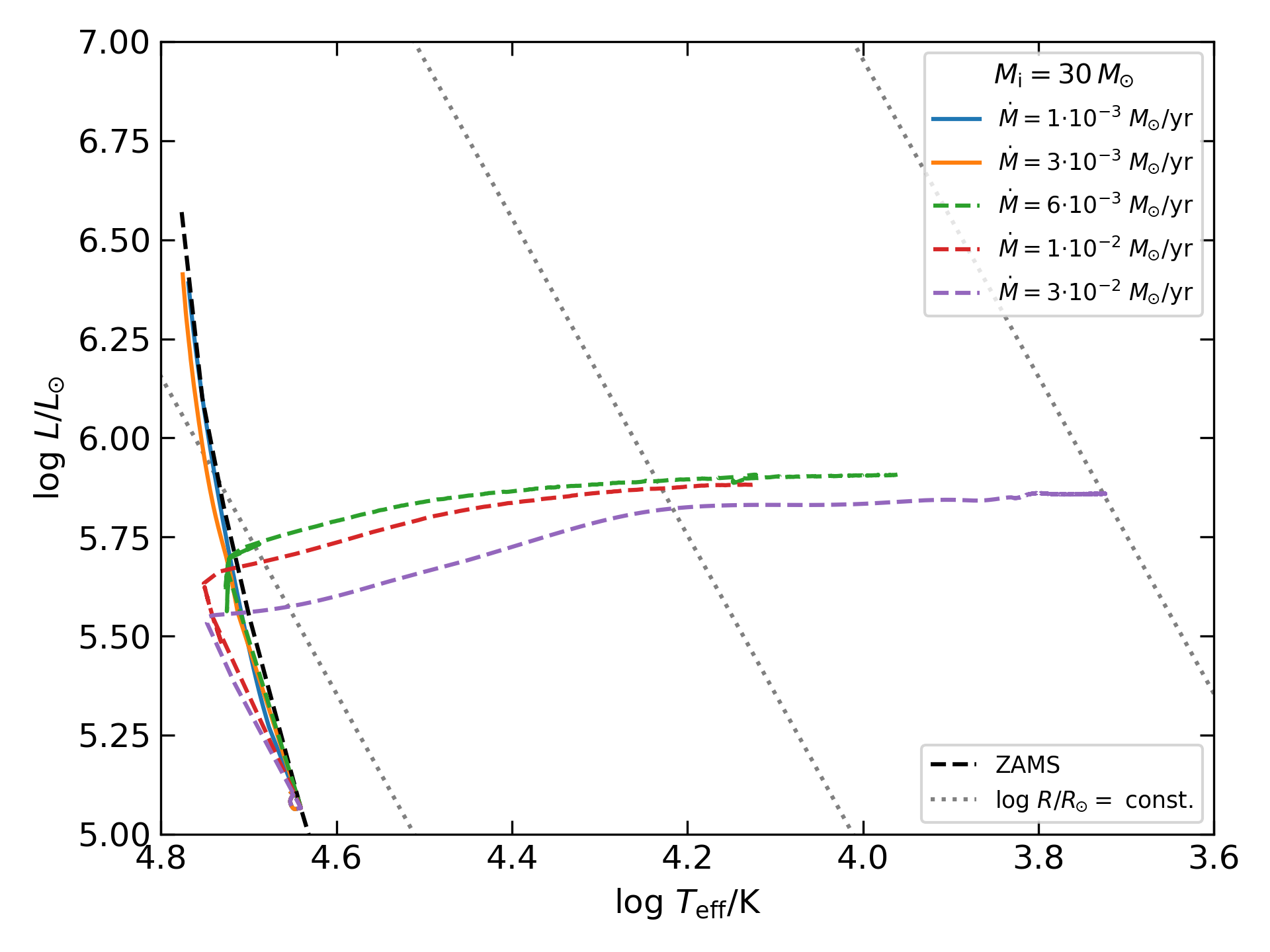}
    \includegraphics[width=0.5\hsize]{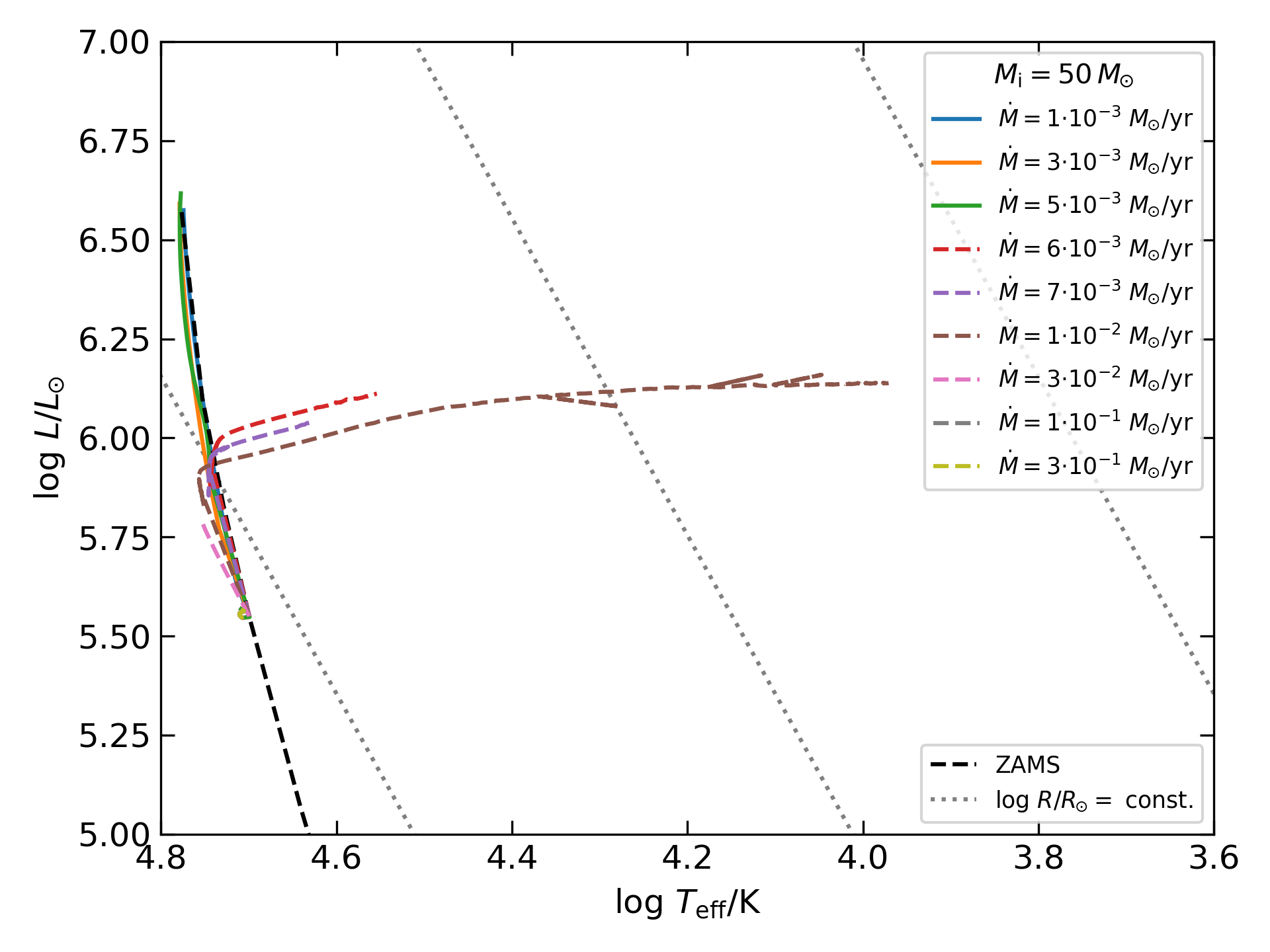}
    \includegraphics[width=0.5\hsize]{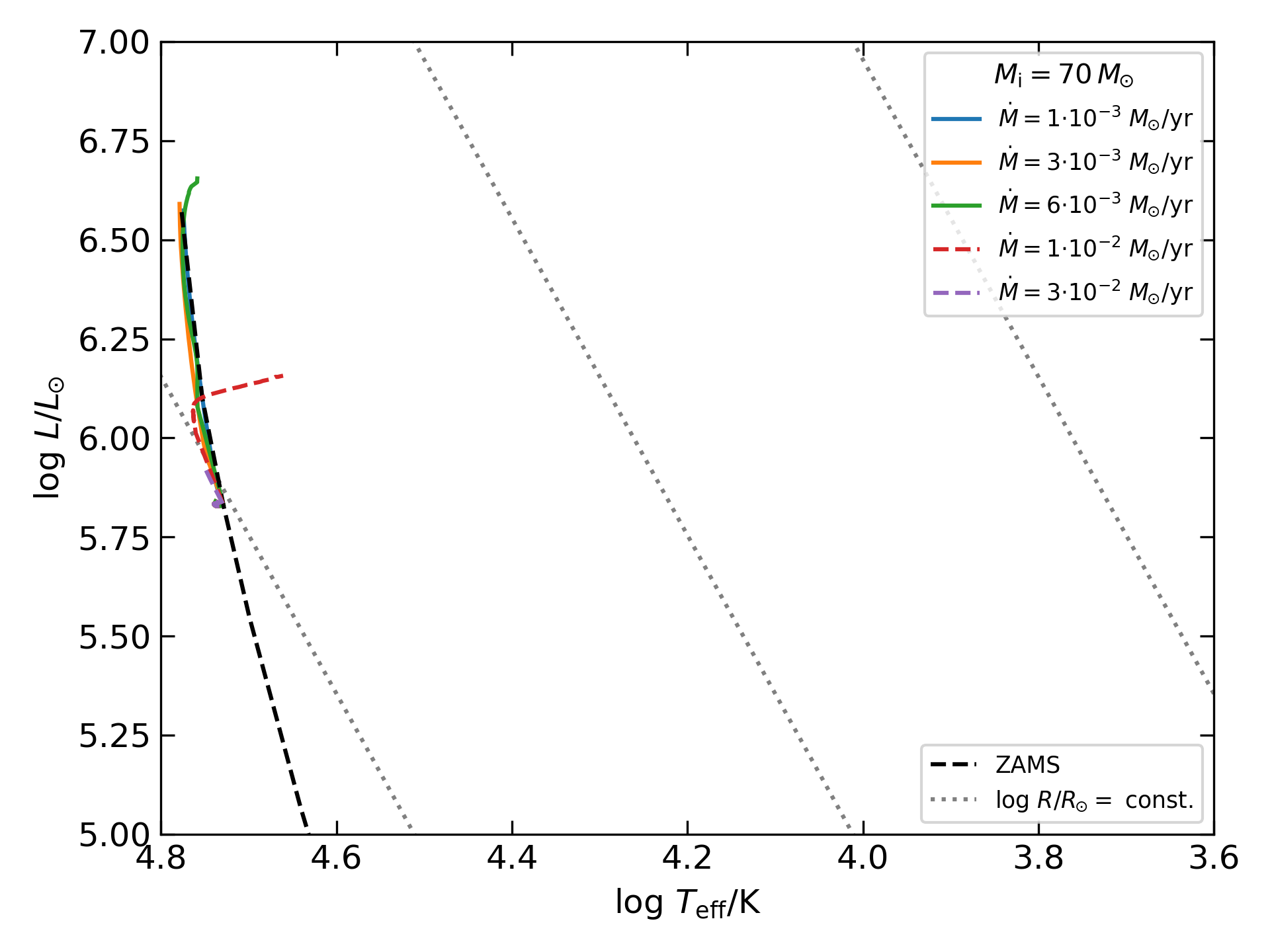}
    \includegraphics[width=0.5\hsize]{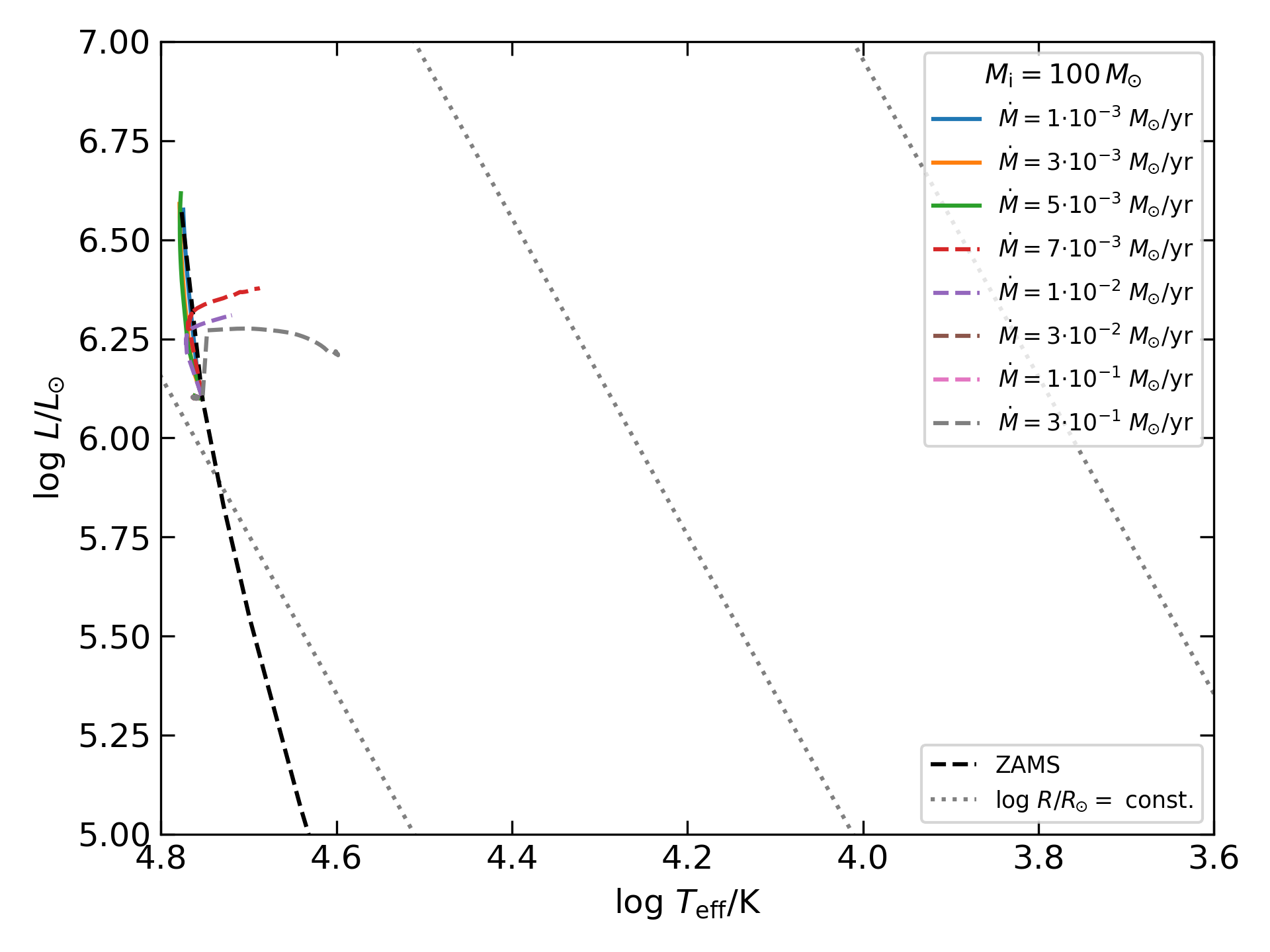}
    \caption{HRDs of our $15\msol$-, $20\msol$-, $30\msol$-, $50\msol$-, $70\msol$-, and $100\msol$-models for different accretion rates (indicated by colour) together with the final mass of the models.}
    \label{fig:hrd2}
\end{figure*}

\newpage\newpage
\section{Boundaries of contact/L2 overflow avoidance regions}
\begin{figure*}[h]
    \includegraphics[width=0.5\hsize]{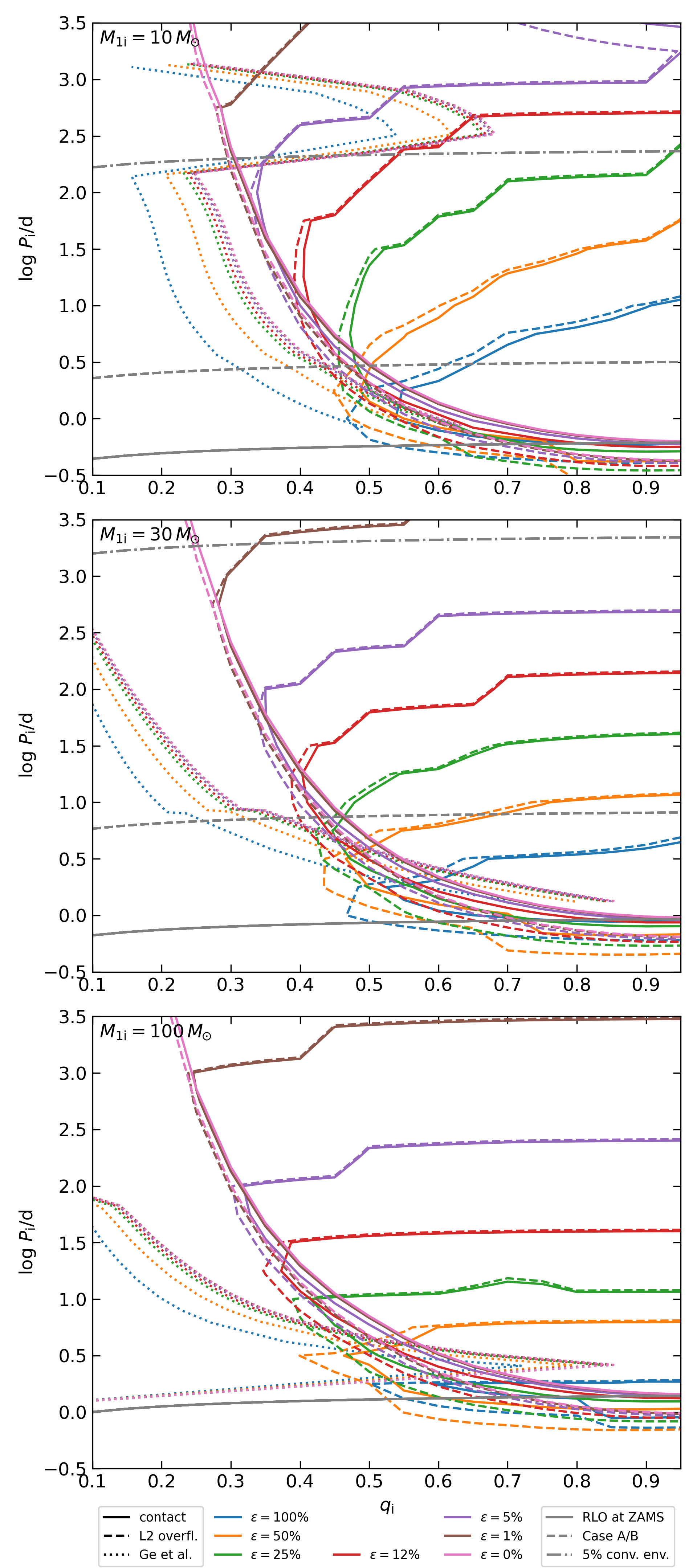}
    \includegraphics[width=0.5\hsize]{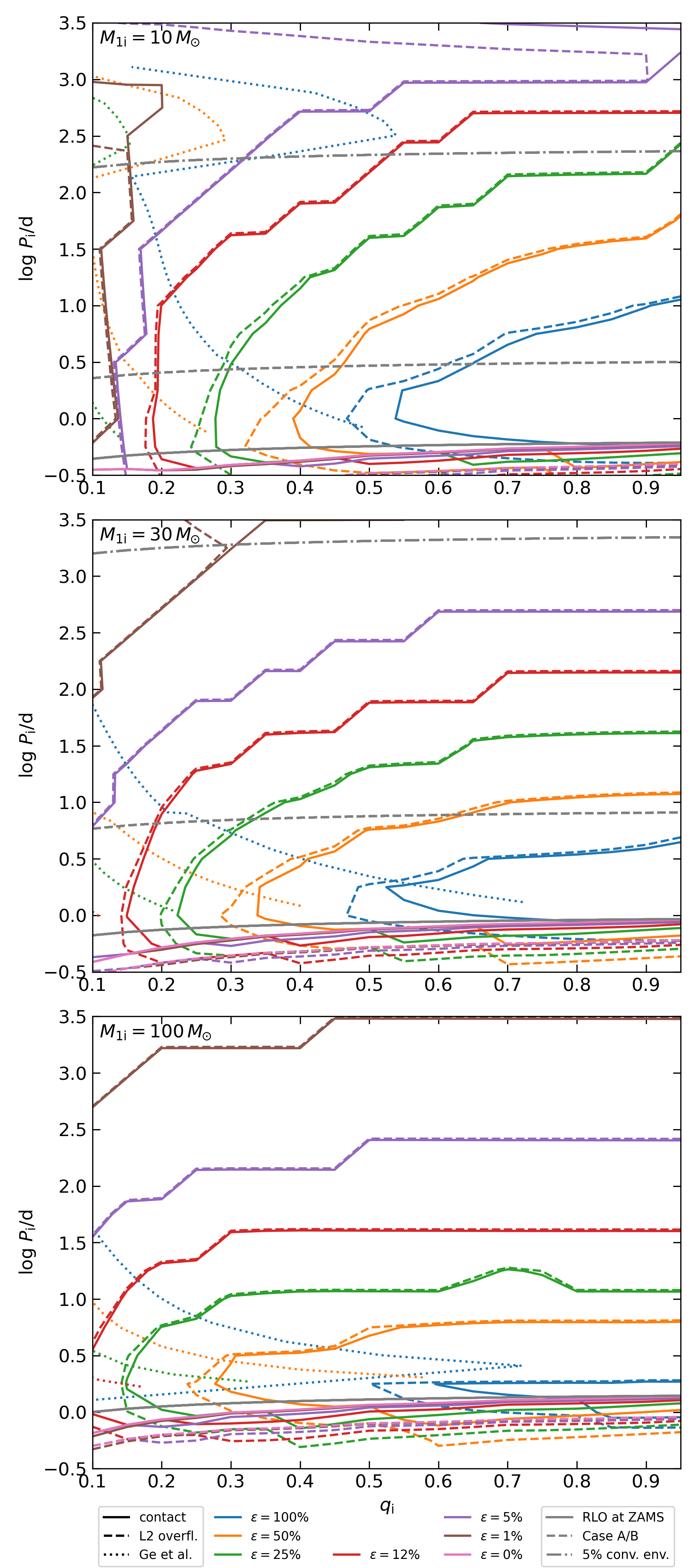}
    \caption{Same as Fig.~\ref{fig:b1}, but the ejected material carries double the donor's specific orbital angular momentum (i.e. $\alpha=\eta=0$, $\beta=-\eps$, $B=2$) or no angular momentum ($A=B=H=0$, right).}
    \label{fig:b20}
\end{figure*}

\begin{figure*}[h]
    \includegraphics[width=0.5\hsize]{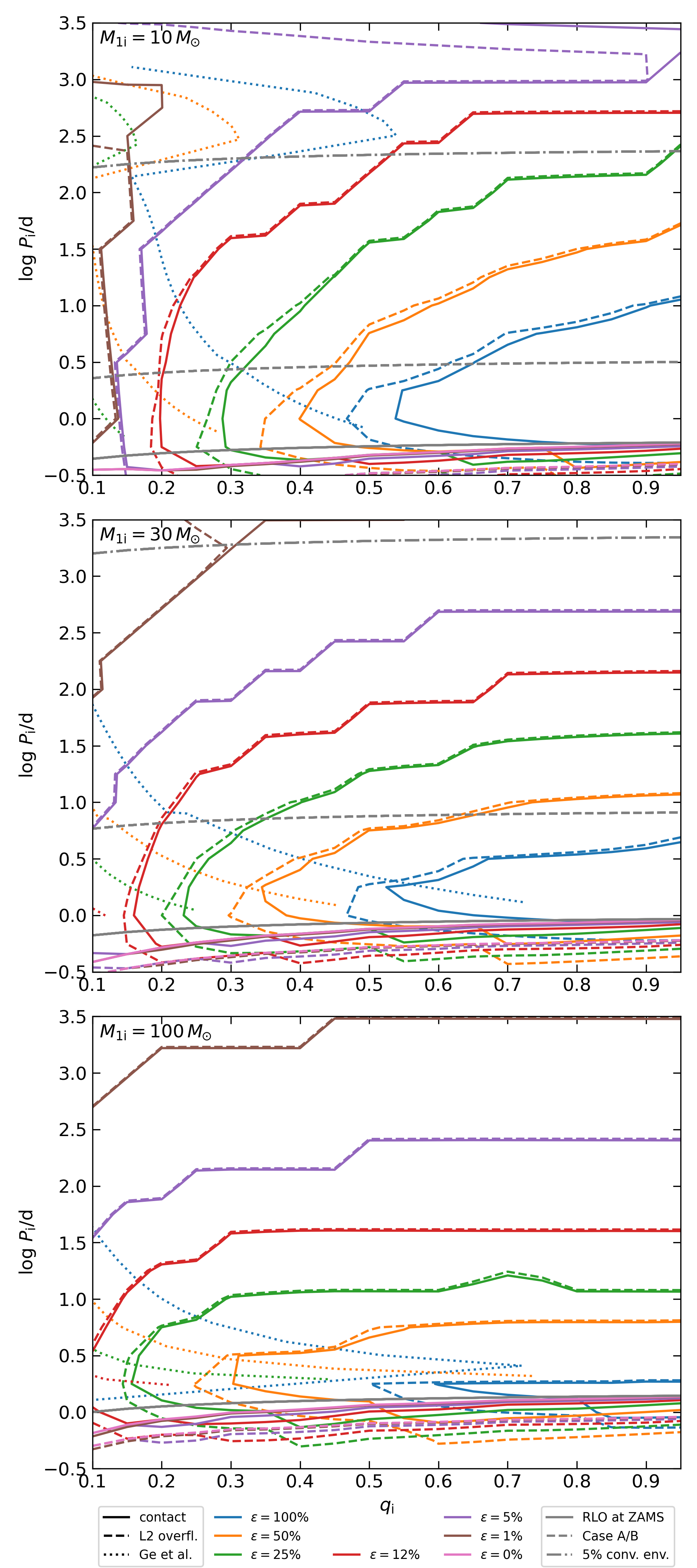}
    \includegraphics[width=0.5\hsize]{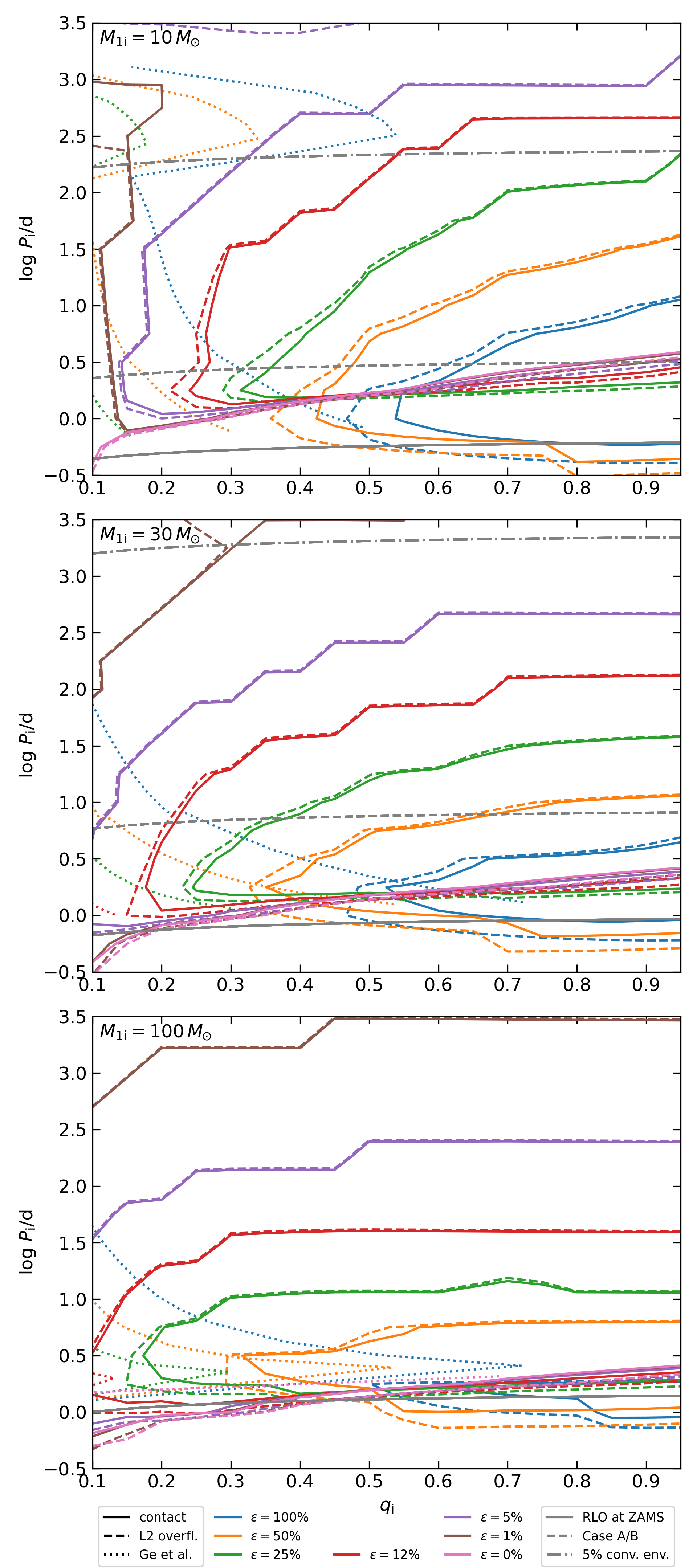}
    \caption{Same as Fig.~\ref{fig:b1}, but the ejected material carries single (i.e. $A=1$, left) or double (i.e. $A=2$, right) the donor's specific orbital angular momentum ($\beta=\eta=0$, $\alpha=-\eps$).}
    \label{fig:a12}
\end{figure*}

\begin{figure*}[h]
    \includegraphics[width=0.5\hsize]{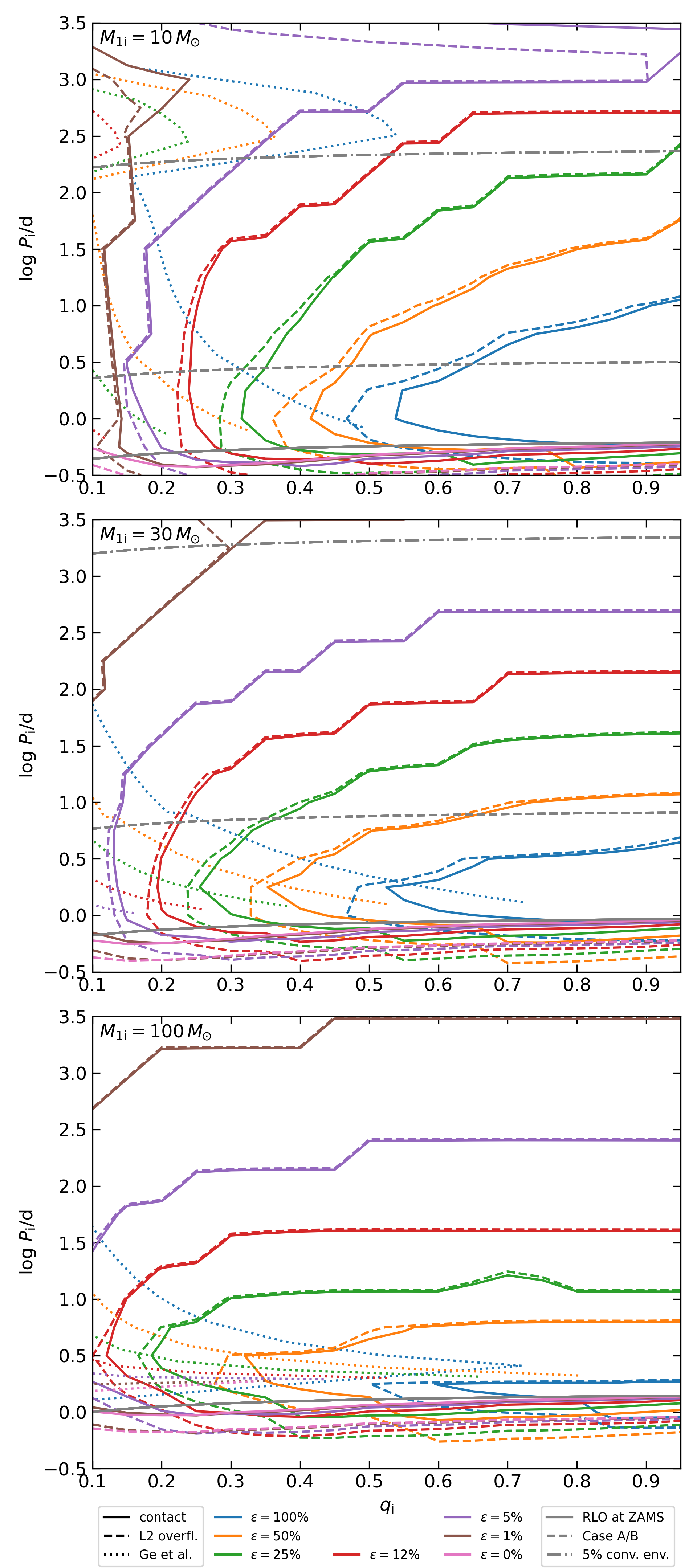}
    \includegraphics[width=0.5\hsize]{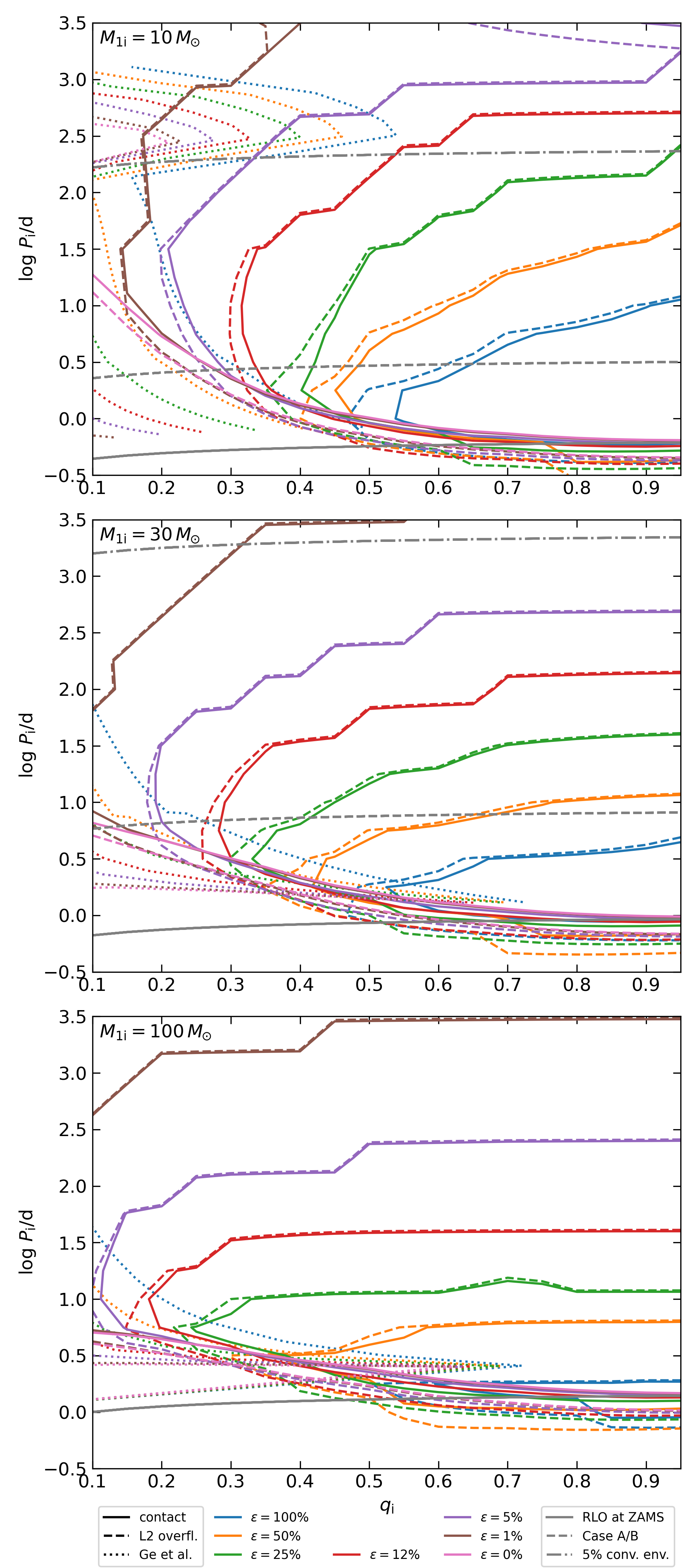}
    \caption{Same as Fig.~\ref{fig:b1}, but the ejected material carries single (i.e. $H=1$, left) or double (i.e. $H=2$, right) the system's specific orbital angular momentum ($\alpha=\beta=0$, $\eta=-\eps$).}
    \label{fig:h12}
\end{figure*}

\begin{figure}[h]
    \includegraphics[width=0.5\hsize]{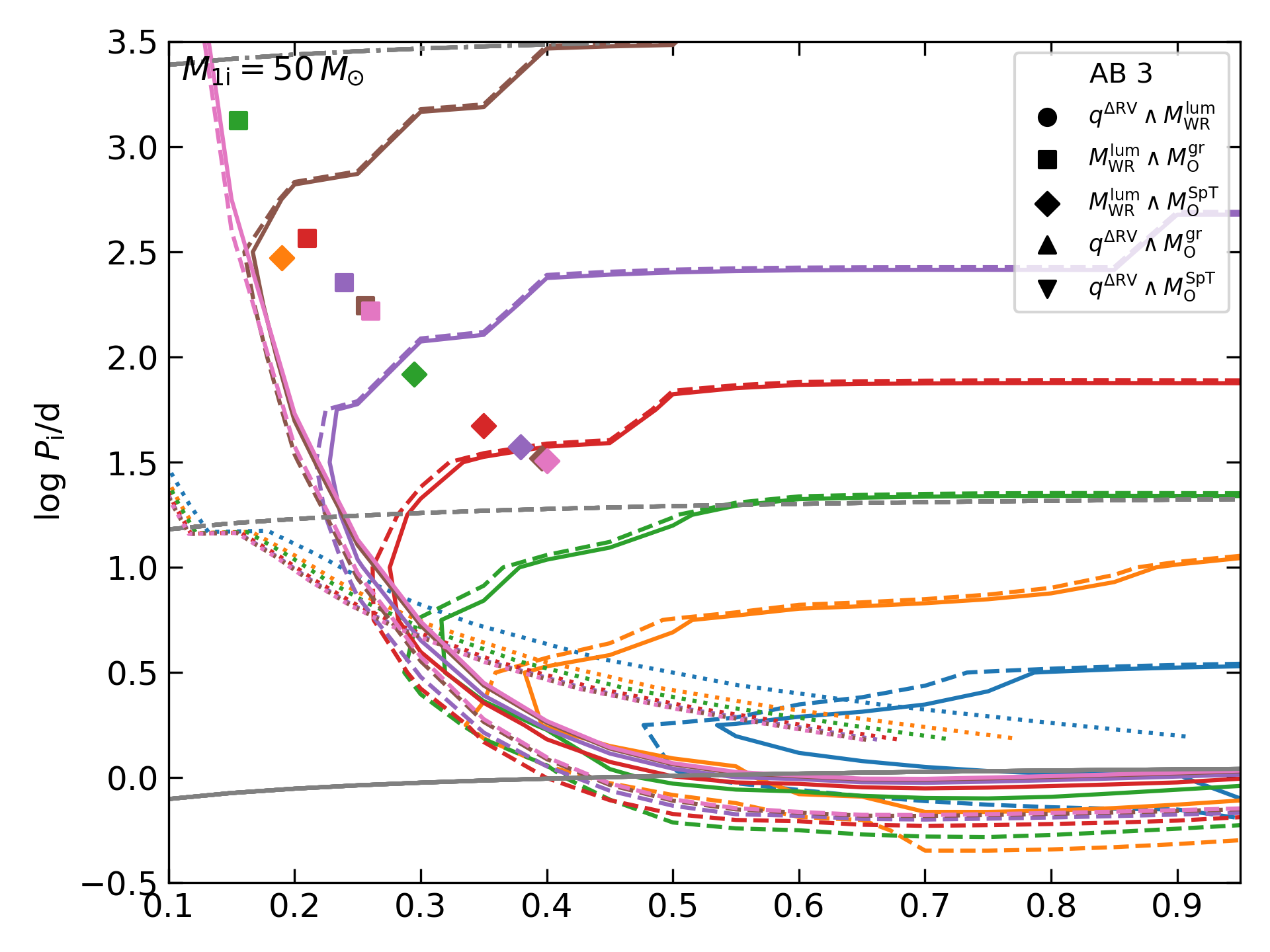}
    \includegraphics[width=0.5\hsize]{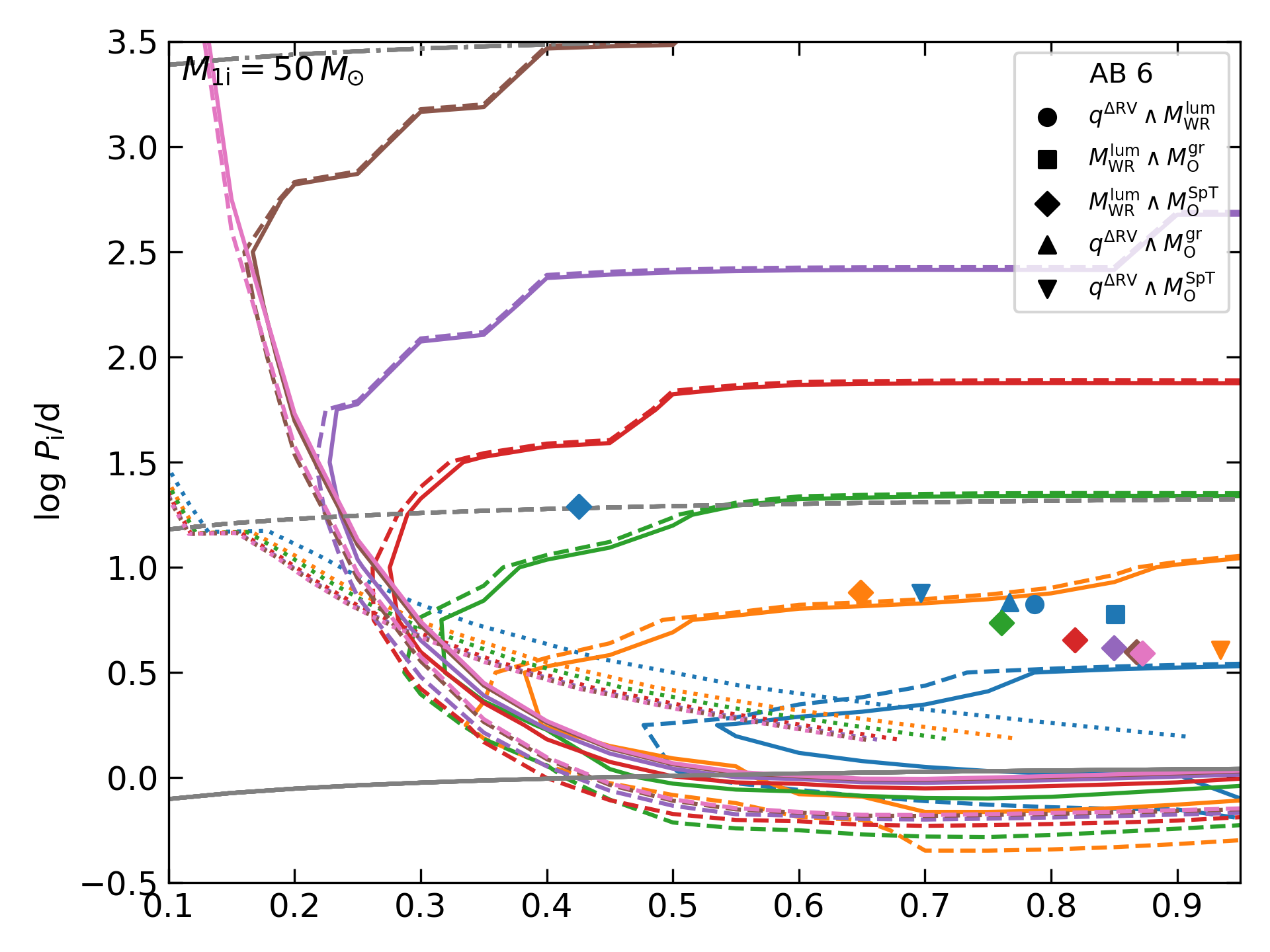}
    \includegraphics[width=0.5\hsize]{pic2/wr_key.png}
    \includegraphics[width=0.5\hsize]{pic2/wr_key.png}
    \begin{center}
    \includegraphics[width=0.5\hsize]{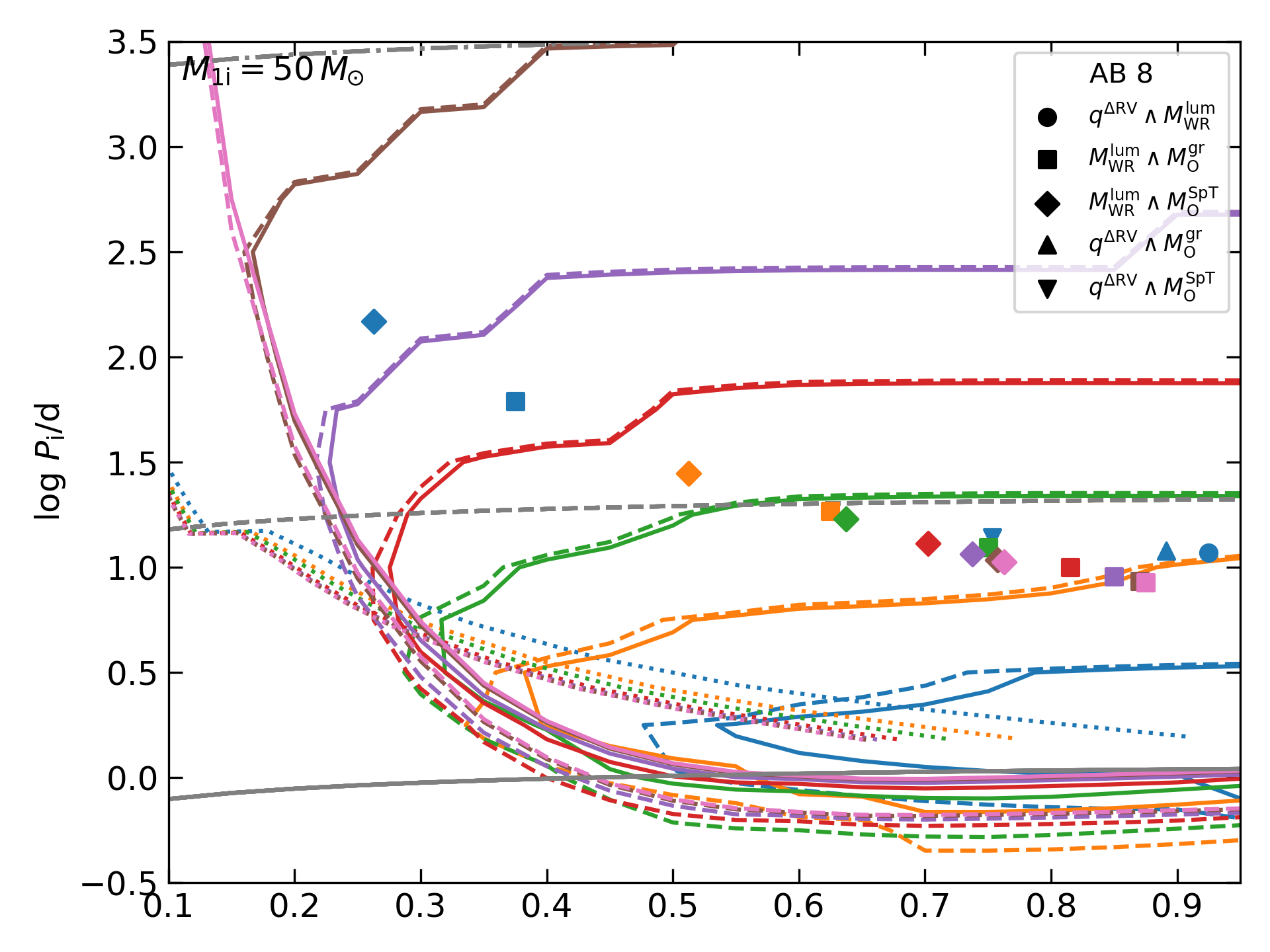}
    \includegraphics[width=0.5\hsize]{pic2/wr_key.png}
    \end{center}
    \caption{Same as Fig.~\ref{fig:wr70}, but for AB\,3 (left), AB\,6 (right), and AB\,8 (bottom). We assumed a WR progenitor mass of $50\msol$, which is a typical number (see Table~\ref{tab:wr}), and since the area of the contact avoiding region varies not so strongly with mass.}
    \label{fig:wr_other}
\end{figure}

\end{document}